\documentclass[%
bibnotes,
amsmath,
amssymb,
pra,
reprint,
superscriptaddress
]{revtex4-2}

\usepackage{graphicx}
\usepackage{dcolumn}
\usepackage{bm}
\usepackage{braket}
\usepackage{mathptmx}
\usepackage[T1]{fontenc}
\usepackage{xcolor}
\usepackage{ulem}

\begin{document}

\title{Resource-efficient high-threshold fault-tolerant quantum computation with weak nonlinear optics}

\author{Kosuke Fukui} 
\affiliation{%
Department of Applied Physics, School of Engineering, The University of Tokyo,\\
7-3-1 Hongo, Bunkyo-ku, Tokyo 113-8656, Japan}
\author{Peter van Loock} 
\affiliation{
Institute of Physics, Johannes Gutenberg-Universität Mainz, Staudingerweg 7, 55128 Mainz, Germany}

\begin{abstract}

Quantum computation with light, compared with other platforms, offers the unique benefit 
of natural high-speed operations at room temperature and large clock rate, but
a big obstacle of photonics is the lack of strong nonlinearities
which also makes loss-tolerant or generally fault-tolerant quantum computation (FTQC) 
complicated in an all-optical setup. Typical current approaches to optical FTQC that aim at building
suitable large multi-qubit cluster states by linearly fusing small elementary resource states would 
still demand either fairly expensive initial resources or rather low loss and error rates.
Here we propose reintroducing weakly nonlinear operations, such as a weak cross-Kerr
interaction, to achieve small initial resource cost and high error thresholds at the same time.
More specifically, we propose an approach to generate a large-scale cluster state by hybridizing 
Gottesman--Kitaev--Preskill (GKP) and single-photon qubits. 
Our approach enables us to implement FTQC based on GKP squeezing of $7.4$ and $8.4$~dB and
a photon loss rate of 1.0 and 5.0~\%, respectively. 
In addition, our scheme has a reduced resource cost, i.e., number of physical qubits/photons
per logical qubit or initial entanglement, compared to high-threshold FTQC with optical GKP qubits
or fusion-based quantum computation with encoded single-photon-qubit states, respectively.
Furthermore, our approach, when assuming very low photon loss, allows to employ GKP squeezing as little as 3.8~dB, 
which cannot be achieved by using GKP qubits alone.
\end{abstract}
\maketitle


Quantum computation (QC) has a great potential to efficiently solve some hard problems for conventional computers~\cite{shor1999polynomial,grover1997quantum}, and realizing QC with light is attractive due to the intrinsic speed and room-temperature operation of optics. 
At the same time, bosonic codes are essential for optical quantum information processing such as fault-tolerant QC (FTQC) and long-distance quantum communication, and employing optical bosonic codes allows to enhance the error tolerance against photon loss.
There exist several types of optical bosonic codes such as low-excitation single-photon codes and the cat~\cite{ralph2003quantum}, binomial~\cite{michael2016new}, and Gottesman--Kitaev--Preskill (GKP)~\cite{gottesman2001encoding} codes that rely upon higher excitation numbers. 
In addition to loss and noise tolerance, another key for practical FTQC in optics is to reduce the resource cost such as the number of qubits and the threshold value required for FTQC.
Thus, many theoretical efforts for practical FTQC in optics have been devoted to realize FTQC with a high photon loss tolerance and low resource cost. 
To realize large-scale QC in an optical setup, measurement-based QC (MBQC) is one of the most promising QC models, where universality can be realized by only adaptive single-qubit measurements on a large-scale cluster state.
Every bosonic code has a unique architecture for MBQC with a characteristic photon loss tolerance. 
Especially, the most advanced approaches to FTQC based on single photons and linear optics exhibit a fairly high loss tolerance~\cite{li2015resource,gimeno2015three,bartolucci2023fusion,pankovich2024high,song2024encoded}, e.g., Ref.~\cite{pankovich2024high} simulated a scheme with loss tolerance as high as $\sim10~\%$.
FTQC based on cat states~\cite{myers2011coherent} requires a large amplitude $\alpha>20$, employing topologically protected MBQC. There also exists FTQC with a small amplitude $\alpha>1.2$, but a photon loss tolerance up to $\sim0.05~\%$~\cite{lund2008fault}.

Among the bosonic codes, the GKP qubit~\cite{gottesman2001encoding} is a promising code for universal and FTQC with continuous variables (CVs)~\cite{pantaleoni2020modular,walshe2020continuous,bourassa2021blueprint,tzitrin2021fault,larsen2021fault,pantaleoni2021subsystem,fukui2022building,schmidt2022quantum,asavanant2023switching,
stafford2023biased,walshe2024linear}.
In particular, the GKP qubit is designed to protect against noise from finite squeezing~\cite{menicucci2014fault} and photon loss~\cite{albert2018performance} during MBQC. 
Furthermore, an important feature is that universal FTQC can be implemented by using only Gaussian operations~\cite{19Baragiola}, once we can prepare a large-scale cluster state for GKP qubits.
In addition, the GKP qubit can achieve the hashing bound of the additive Gaussian noise with a suitable quantum error correcting code~\cite{fukui2017analog,fukui2018high}. 
However, an optical GKP qubit~\cite{konno2024logical} as required for FTQC has not been available experimentally.
Thus, there is a demand for a further improvement of the threshold of the squeezing level for FTQC with GKP qubits. 

Interestingly, an optical hybrid approach~\cite{van2011optical,andersen2015hybrid} combining the cat code with single-photon qubits is also useful for FTQC~\cite{lee2013near,jeong2014generation,lee2015nearly}. Specifically, FTQC based on hybrid photon-cat qubits has been proposed~\cite{omkar2020resource,omkar2021highly,lee2024fault}.
Moreover, sufficient nonlinearities for optical quantum information processing are obtainable using a coherent state with a large amplitude even if the strength of the nonlinear interaction is small, referred to as a weak cross-Kerr (CK) scheme.
There have been many works on quantum information processing with a weak CK scheme, e.g. the controlled-NOT gate between photons~\cite{nemoto2004nearly}, quantum error correction~\cite{yamaguchi2006quantum,myers2007stabilizer}, and QC based on the so-called qubus model~\cite{munro2005weak,spiller2006quantum,van2008hybrid}. 
Although it is difficult to obtain large nonlinearities in optics, the nonlinearity required for the weak CK scheme does not need to be so large. Such a nonlinear interaction is expected to be available with state-of-the-art technology~\cite{venkataraman2013phase,feizpour2015observation}.

The main goal of this work is to provide a novel approach for resource-efficient FTQC by combining GKP qubits with single photons.
In our scheme, we may use a CK interaction to implement the two-mode gate between the GKP and the single-photon qubits.
In this model, MBQC with GKP qubits is implemented after the construction of a large-scale cluster state through Bell measurements~(BMs) between hybrid bosonic qubits via, for instance, a weak CK interaction.
As a result, our scheme overcomes limitations for optical FTQC with CV alone~\cite{note3} and achieves a high efficiency regarding resource costs for optical FTQC.  
Furthermore, we show that our scheme can also achieve an ultra-high squeezing noise threshold under the assumption of very low photon loss, which even exceeds the ultimately obtainable error-tolerance of a phenomenological noise model for GKP qubits alone. 
While we employ a weak and lossy CK interaction to present the concept of our nonlinear optical approach to FTQC, the scheme does not depend on physically available Kerr nonlinearities, which may be problematic when applied to single-photon qubits~\cite{shapiro2006single,gea2010impossibility}. The nonlinear gates and measurements may also be based upon elementary cubic interactions~\cite{yanagimoto2023mesoscopic} or their decompositions~\cite{budinger2024all}.

{\it The GKP qubit.}---
Gottesman, Kitaev, and Preskill proposed a method to encode a qubit in an oscillator's $q$ (position) and $p$ (momentum) quadratures to correct errors caused by a small deviation in the $q$ and $p$ quadratures~\cite{gottesman2001encoding}. 
The basis of the GKP qubit with finite squeezing is composed of a series of Gaussian peaks of width $\sigma$ and separation $\sqrt{\pi}$ embedded in a larger Gaussian envelope of width 1/$\sigma$. 
The codeword 0 and 1 states $\ket {\widetilde{l}}$ $(l=0,1)$ are defined as  
\begin{eqnarray}
\ket {\widetilde{l}} &\propto &  \sum_{t=- \infty}^{\infty} \int dx e^{-\frac{[(2t+l)\sqrt{\pi}]^2}{2(1/\sigma^2)}}  
{e}^{-\frac{(x-(2t+l)\sqrt{\pi})^2}{2\sigma^2}}\ket{x}_q .     \label{gkp}
\end{eqnarray}
The squeezing level $s$ is defined by $s=-10{\rm log}_{10}2{\sigma}^2$.
There is a finite probability of misidentifying $\ket {\widetilde{0}}$ as $\ket {\widetilde{1}}$, and vice versa. 
Provided the magnitude of the true deviation is more than $\sqrt{\pi}/2$ from the peak value, the decision of the bit value is incorrect. 
The probability of misidentifying the bit value,~$E(\sigma^2)$, is calculated by
\begin{equation}
E(\sigma^2) = 1-\int_{\frac{-\sqrt{\pi}}{2}}^{\frac{\sqrt{\pi}}{2}} dx \frac{1}{\sqrt{2\pi {\sigma} ^2}} e^{-\frac{x^2}{2{\sigma} ^2}}.
\label{err}
\end{equation}

{\it Weak CK interaction scheme.}---
The Hamiltonian for a CK interaction, $\hat{H}_{\rm CK}$, is described as
$
\hat{H}_{\rm CK}=\hbar \chi \hat{a}^{\dag}\hat{a}\hat{b}^{\dag}\hat{b}
$
with the interaction strength $\chi$, where $\hat{a}(\hat{b})$ and $\hat{a}^\dag(\hat{b}^\dag)$ are the annihilation and creation operators for mode a (b), respectively.
We employ the technique with a weak CK interaction between two optical modes.
We consider the interaction between a single photon and a coherent state, assuming the photon state $\ket{\phi}=(\ket{0}+\ket{1})/\sqrt{2}$ and a coherent state $\ket{\alpha}$ $( \alpha \in  \mathbb{R})$ with $\alpha \gg 1$.
The interaction with interaction time $t$ transforms the two states as
$
 {e}^{-\frac{it}{\hbar}{\hat{H}_{\rm CK}}}\ket{\phi}\ket{\alpha}\mapsto
(\ket{0} \ket{\alpha}+\ket{1} \ket{\alpha {e}^{-i\chi t}})/\sqrt{2}.
$
Here we assume that the phase shift per photon, $\theta=\chi t$, is small, e.g. $\theta=10^{-5}$ is assumed in Refs.~\cite{nemoto2004nearly,munro2005weak}. Then $\ket{\alpha {e}^{-i\theta}}$ is approximated to $\ket{\alpha -i\alpha\theta}$.
To see the effect depending on a single photon,
we redefine $\ket{\alpha}$ and $\ket{\alpha -i \alpha\theta}$ as $\ket{0}$ and $\ket{-i \alpha\theta}$, respectively.
Then $ {e}^{-\frac{it}{\hbar}{\hat{H}_{\rm CK}}}\ket{\phi}\ket{\alpha}$ can be rewritten as
$
\ket{0} \ket{0}+\ket{1} \ket{ -i\alpha\theta}
$,
which means that a large $\alpha$ provides a sufficient interaction to realize a large displacement in the quadrature even for small $\theta$.
In this work, we use an appropriate amplitude of the coherent state so that the displacement in the $q$ quadrature is equal to $\sqrt{\pi}$ to implement the bit-flip operation for the GKP qubit.

{\it Hybrid GKP-photon qubit.}---
We consider a weak CK interaction between the GKP qubit $\ket {\widetilde{0}}$ and photon $\ket{\phi}=(\ket{0}+\ket{1})/\sqrt{2}$, as shown in Fig.~\ref{fig2}(a).
In our scheme, we employ the operation as given by
$
\hat{U}_{\rm CX}=\hat{D}(-\alpha){e}^{-\frac{i t}{\hbar}{\hat{H}_{\rm CK}}}\hat{D}(\alpha),
$
where $\hat{D}(\alpha)$ represents a displacement operation with amplitude $\alpha$ in the $p$ quadrature.
Here $\hat{U}_{\rm CX}$ acts as the controlled-not gate between a single photon and the GKP qubit, where the single photon and the GKP qubit are control and target qubits, respectively.
Then $\hat{U}_{\rm CX}$ transforms the two states to
$
\hat{U}_{\rm CK}\ket{\phi}\ket {\widetilde{0}}\mapsto (\ket{0} \ket{\widetilde{0}}+\ket{1}\ket{\widetilde{1}^\ast})/\sqrt{2}
$
where the approximated logical $0$ and $1$ state, $\ket {\widetilde{0}^{\ast}}$ and $\ket {\widetilde{1}^{\ast}}$, are defined as $\hat{U}_{\rm CX}\ket {1}\ket {\widetilde{0}}=\ket {1}\ket{\widetilde{1}^\ast}$ and $\hat{U}_{\rm CX}\ket {1}\ket {\widetilde{1}}=\ket {1}\ket{\widetilde{0}^\ast}$, respectively.
We note that the infidelity between $\ket{\widetilde{0}}(\ket{\widetilde{1}})$ and $\ket{\widetilde{0}^{\ast}}(\ket{\widetilde{1}^{\ast}})$ is negligibly small for the squeezing level considered in this work and $\alpha_0\sim 3\times 10^{5}$ as assumed in Refs.~\cite{nemoto2004nearly,munro2005weak}.
 (See the supplemental material A for details on $\hat{U}_{\rm CX}$ and the fidelity.)

\begin{figure*}[t]
\centering \includegraphics[angle=0, width=2.0\columnwidth]{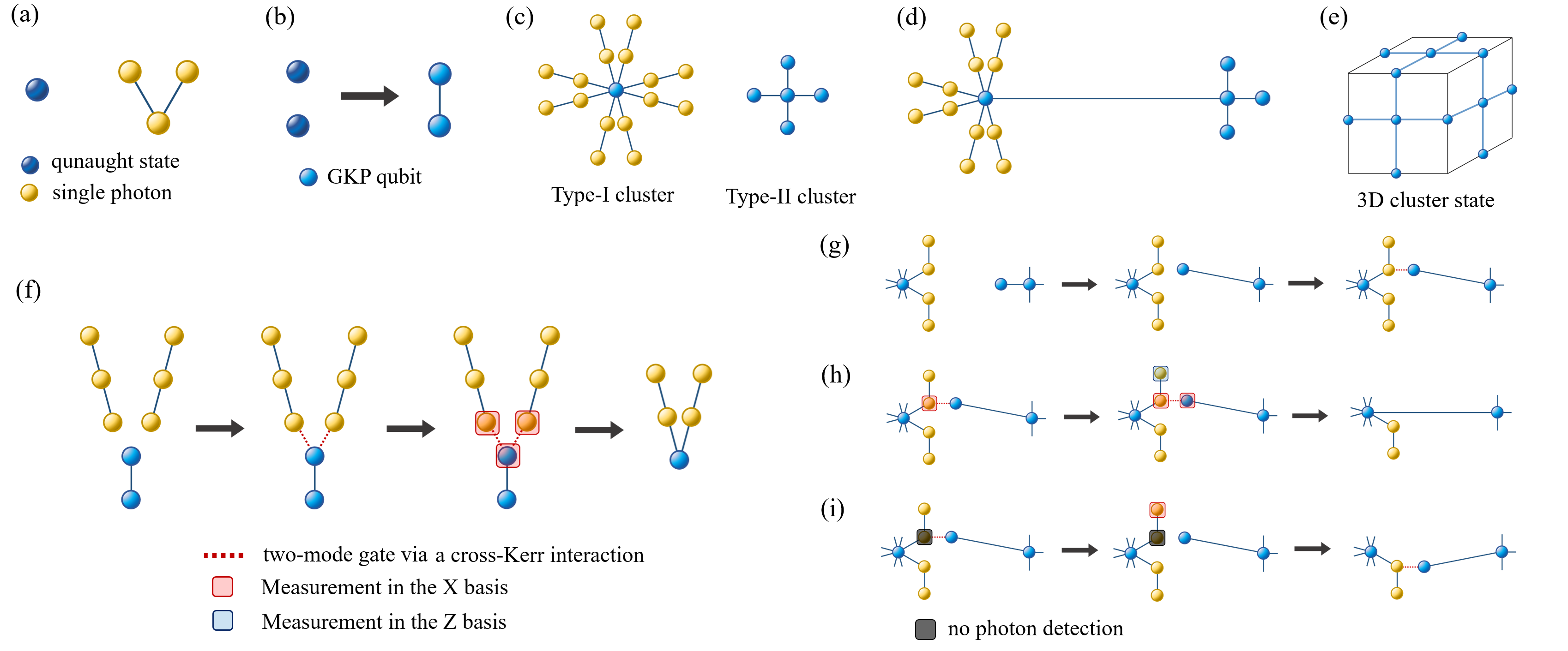} 
     \setlength\abovecaptionskip{0pt}
\caption{Schematic view of our scheme. 
(a) The elemental quantum states. 
(b) The entangled GKP qubits generated from two qunaught states. The Bell state of the GKP qubits is prepared by a beam-splitter coupling between the two qunaught states. The cluster state of the two GKP qubits is prepared by a Fourier transform acting on one of the qubits in the Bell state. (c) The type-I and-II cluster states. The type-I cluster is composed of a single GKP qubit and 16 single photons. The type-II cluster is composed of 5 GKP qubits.
(d) The entanglement generation between two node qubits via two types of small-scale clusters, which is used to generate the 3D cluster state. (e) The 3D cluster state used for CV-FTQC. (f) An example for the generation of the hybrid cluster state consisting of a single GKP qubit and 4 photons. 
(g)--(i) The entanglement generation between two node qubits. (g) A single GKP qubit sent from the type-II cluster is entangled with photons of the type-I cluster.
(h) When the photon measurement succeeds, we keep the entanglement. (i) When the photon measurement fails, we discard entanglement via the photon measurement in the $Z$ basis, and then we try entanglement generation using another photon of the type-I cluster.
}
\label{fig1}
\end{figure*}

\begin{figure}[t]
\centering \includegraphics[angle=0, width=0.8\columnwidth]{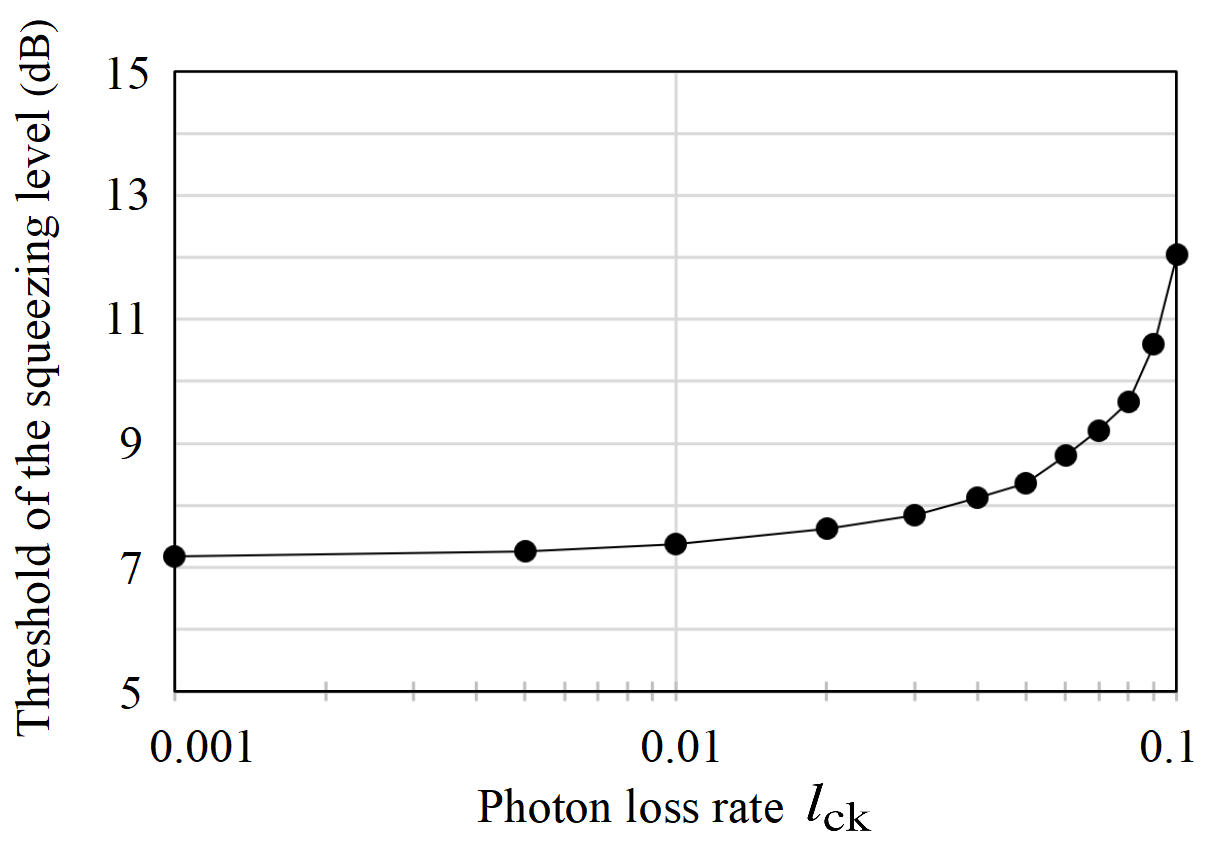} 
     \setlength\abovecaptionskip{0pt}
  \caption{
The threshold of the squeezing level for CV-FTQC as a function of $l_{\rm ck}$ with $l_{ \rm det}=0.001$, where $l_{\rm ck}$ and $l_{ \rm det}$ are photon loss for the CK interaction and the photon detector, respectively.
}
\label{fig2}
\end{figure}

{\it FTQC model.}---
Figure~\ref{fig1} shows our scheme consisting of three concepts; (1) Deterministic resource of the elemental quantum states, e.g. the cluster state of photons and the qunaught state, as shown in Fig.~\ref{fig1}(a). The two qunaught states are used to prepare the entangled GKP qubits, as shown in Fig.~\ref{fig1}(b).
(2) Probabilistic preparation of the two types of small-scale clusters from the elemental quantum states via the weak CK scheme, as shown in Fig.~\ref{fig1}(c).
(3) Deterministic large-scale cluster state preparation from the small-scale clusters via the weak CK scheme, as shown in Fig.~\ref{fig1}(d). 
Then, we perform topologically protected MBQC on the generated large-scale cluster state, e.g. the 3D cluster state consisting of the GKP qubits, as shown in Fig.~\ref{fig1}(e). 

 {\it The cluster state construction.}---We describe the state preparation for the proposed FTQC model in more detail.
 As an example of the construction, we consider making a hybrid cluster state consisting of a single GKP qubit and 4 photons via the CK interaction scheme from a GKP qubit pair and two single-photon GHZ states, as described in Fig.~\ref{fig1}(f).
For the preparation of the hybrid cluster in Fig.~\ref{fig1}(f), we start to prepare the entangled GKP qubits from two qunaught states deterministically, as described in Fig.~\ref{fig1}(b).
Then, we implement the entanglement generation between the single GKP qubit and one photon of each GHZ state via the CK interaction.
After the interaction, we measure the GKP qubit and the two corresponding photons in the $X$ basis, where the GKP qubit and photons are measured by using homodyne measurement and photon detectors, respectively.
We refer to this entanglement operation using the CK interaction and $X$ basis measurements as a hybrid BM.
After the hybrid BM with an appropriate feedforward operation, a hybrid cluster of 4 photons and the single GKP qubit is prepared.
By using the hybrid BM, we prepare two types of small-scale clusters, as shown in Fig.~\ref{fig1}(c).
 (See the supplemental material B for details on the small-scale cluster construction.)
To reduce the effect of photon loss in the hybrid BM, we use two postselected schemes.
One is the highly reliable measurement scheme used to decrease the measurement error probability of the GKP qubit, as introduced in Ref.~\cite{fukui2018high}, to prepare a reliable large-scale cluster state.
Another one is based on discarding the case where a single photon is not counted in the photon measurement, which allows to detect photon loss events and reduce photon loss errors.
The small-scale clusters are successfully generated if all postselections are successful.
Consequently, the hybrid BM with postselections allows to make loss-tolerant and reliable states during large-scale cluster construction.
After the preparation of two types of cluster states, we construct a large-scale cluster state.  In this step, we partially employ postselection since the large-scale cluster must be prepared deterministically to perform FTQC deterministically.
Once the small clusters are prepared in the nodes corresponding to each qubit in the 3D cluster state, several qubits are sent to neighboring nodes to generate entanglement between these nodes, as shown in Fig.~\ref{fig1}(d).
We perform the entanglement generation between neighboring nodes via the hybrid BM with partial use of postselection, as shown in Fig.~\ref{fig1}(g)--(i).
We first try to generate entanglement between a GKP qubit and a photon in the upper mode, as shown in Fig.~\ref{fig1}(g).
When the photon in the upper mode is detected, we measure an ancilla photon in the upper mode, as shown in Fig.~\ref{fig1}(h), and qubits in the lower mode are also measured in the $Z$ basis to remove the entanglement.
When the photon detection fails in the upper mode, we measure the ancilla photon in the upper mode to remove the entanglement, and we try another hybrid BM in a lower mode, as shown in Fig.~\ref{fig1}(i).
Thanks to two tries of the hybrid BM,  we can discard some loss events.
Consequently, entanglement between neighboring-node qubits in the 3D cluster state is generated in a reliable manner, as shown in Fig.~\ref{fig1}(d).

{\it Error model.}---For the calculation of the threshold for CV-FTQC, we calculate the measurement error probability of the qubits in the 3D cluster state.
The measurement error comes from a node GKP qubit itself and the accumulated errors during the preparation of the 3D cluster state.
The accumulated errors are the measurement errors from each qubit, e.g., the measurement errors of the GKP qubit and single photon in the hybrid BM, which accumulate into a node qubit after the generation of the 3D cluster state.
For the GKP qubit, the measurement error is calculated by Eq.~(2), which depends on the squeezing level $s=-10{\rm log}_{10}2{\sigma}^2$.
Then, the effect of loss on the GKP qubit is considered as an increase of the variance, $\sigma^2$.
After photon loss, the variance changes as,
$
{\sigma^{2}}$ $\mapsto$ ${\eta} \sigma^{2} +({1-{\eta}})/{2}
$,
where the photon loss rate is given by 1- $\eta$.
In the measurement including photon loss, the outcome from the homodyne measurement is multiplied by $1/\sqrt{\eta}$ in the classical post-processing so that the peaks of the GKP qubit are fixed at integer multiples of $\sqrt{\pi}$ due to the GKP codewords.
Consequently, the probability of misidentifying the bit value after photon loss is calculated by~$E({\sigma'}^2)$ with ${{\sigma'}^{2}}$ = $\sigma^{2}+({1-{\eta}})/{2{\eta}}$.
For a single photon, the measurement errors come from the photon loss and the depolarization error, where the measurement error probabilities of photon loss and depolarization error are assumed to be 0.5 and $4.2\times 10^{-5}$~\cite{azuma2015all}, respectively.
In this work, we consider a model with two main sources of photon loss.
First, we consider photon loss in the CK interaction as photon loss before the CK interaction, which occurs on the single photon and the GKP qubit, assuming the photon loss to be $l_{\rm ck}$. For the GKP qubit, we incorporate this effect as $\eta=l_{\rm ck}$.
Second, we assume a photon loss of $l_{\rm det}$ in the photon detectors.

{\it Threshold of the squeezing level.}---
We numerically investigate the threshold for topologically protected MBQC using the generated 3D cluster state.
In the decoding, we use the minimum-weight perfect-matching algorithm~\cite{edmonds1965paths, kolmogorov2009blossom} with analog quantum error correction with the GKP qubits~\cite{fukui2017analog}.
Figure~\ref{fig2} shows the logical error probabilities as a function of the squeezing level.
When the error is less than a certain value, i.e. the threshold value, the logical error probability decreases as the code size increases.
The numerical calculations show that our method can achieve a required squeezing level~$\sim 7.4$~dB with photon loss~1.0~\%.
In addition, our scheme realizes FTQC with  $\sim$7.2~dB and $\sim$8.4~dB with photon loss $l_{\rm ck}=$0.1~\% and 5.0~\%, respectively~\cite{note4}.

{\it Resource cost.}---
We examine the resource cost for CV-FTQC with our scheme by calculating the required number of qubits to construct each node qubit in the 3D cluster state. 
We count an average qubit number for the two types of small-scale clusters needed to construct each node qubit.
 (See the supplemental material C for details on the calculation of the resource cost.) 
We obtained the average of the required qubits as $\sim256$ and $\sim86$ to realize a squeezing level threshold of 7.4~dB with a photon loss rate of 1.0~\% and 5.0~\%, respectively.
Here, we compare the required number of qubits for FTQC with GKP qubits. In Refs.~\cite{fukui2018high, fukui2023high}, FTQC can be realized under moderate squeezing levels, such as 8.1~dB and 9.8~dB, respectively. However, a large number of qubits is required in both FTQC schemes due to the use of postselection.
Specifically, Ref.~\cite{fukui2018high} requires $9.2 \times 10^6$ qubits at a threshold squeezing level of 9.8~dB, while our scheme requires much fewer qubits. We note that Ref.~\cite{tzitrin2021fault} requires a moderate number of qubits for CV-FTQC due to the absence of postselection; however, the required squeezing level is higher than in our scheme, e.g., 10.1 dB without photon loss.

In addition, our approach may have an advantage in resource cost over fusion-based QC with encoded single-photon-qubit states.
Specifically, Ref.~\cite{pankovich2024high} proposed FT universal cluster-state generation with a loss tolerance as high as $\sim10~\%$ and a resource cost of about 100 qubits per node. However, these photonic qubits must be initially entangled in an elementary resource state whose generation efficiency and fidelity, though independent of the size of the final cluster state, will also contribute to the overall performance and practicality of this approach to FTQC. 
Furthermore, FTQC based on hybrid photon-cat qubits can be resource-efficient~\cite{lee2024fault}, e.g., with a resource cost of about 100 Bell pairs of the hybrid photon-cat qubits to generate small-scale cluster states for the entanglement generation between node qubits.
However, in this case, the loss threshold is less than 1.0~\% at most.
Thus, our scheme offers resource-efficient FTQC with both a lower squeezing level and reduced resource cost.

{\it FTQC model under negligible photon loss.}---
We finally observe that a hybrid bosonic scheme with very low photon loss enables us to achieve the threshold corresponding to the bound of the phenomenological error model, 3.5 dB~\cite{note1}. 
In principle, the threshold for the phenomenological model represents a lower bound for CV-FTQC under the assumption that displacement errors occurring in one GKP qubit do not propagate to other qubits.
This means there is no room for further improvement in the threshold value of squeezing when one begins to generate cluster states from GKP qubits alone for finite squeezing.
We shall briefly describe the model under the assumption of negligible photon loss (see Suppl. Mat. D for details).
Considering negligible photon loss, we assume deterministic photon detection. To construct a 3D cluster state of GKP qubits, each node GKP qubit shares photon Bell pairs with its four neighboring nodes. Then, we implement the CK interaction between each GKP qubit and the shared photons. After measurement of the single photons in the $X$ basis, we obtain the 3D cluster.
In the numerical simulation, we found a threshold value of 3.8~dB~\cite{note2}.
Thus, these results indicate that our approach has the potential to achieve the bound of the squeezing level within the phenomenological error model, although the basic loss assumption may not be practical.

{\it Discussion and conclusion.}---
We have developed resource-efficient FTQC using hybrid GKP-photon qubit states and employing a weak-cross Kerr interaction to implement the entanglement generation between the GKP and single-photon qubits.
The numerical calculations showed that the threshold of the GKP squeezing can be considerably improved to 7.4~dB with a photon loss of 1.0~\%.
In addition, the threshold with very low photon loss can approach 3.8~dB, which corresponds to the threshold of a phenomenological error model for the GKP qubit, which cannot be achieved by using GKP qubits alone. 
Furthermore, the number of qubits required for FTQC is significantly reduced compared to other optical architectures for high-threshold FTQC with a required squeezing level of 8.1~dB.
Thus, our scheme offers a novel approach to achieve CV-FTQC with high threshold and a moderate number of qubits.
Therefore, the proposed approach overcomes the limitations of optical FTQC with CV alone.
In addition, our scheme can be potentially applied to quantum communication~\cite{bennett2014quantum,briegel1998quantum,kimble2008quantum} as well 
thanks to its robustness against photon loss, along similar lines to recent research on quantum communication using GKP states~\cite{fukui2021all,rozpkedek2021quantum,rozpkedek2023all,schmidt2024error,haussler2024quantum}.
Our approach here, when applied to GKP-based quantum communication protocols, is also expected to reduce the resource cost there.
For the preparation of the elementary states, there have been various proposals to generate the optical GKP qubit~\cite{motes2017encoding,weigand2018generating,arrazola2019machine,eaton2019non,su2019conversion,fukui2022efficient,fukui2022generating,hastrup2022protocol,takase2023gottesman,yanagimoto2023quantum,budinger2024all,pizzimenti2024optical}.
In future works, we will analyze FTQC including the imperfection during generation of elementary quantum states.
An interesting future topic could be to explore whether our approach would still be possible without weak nonlinear gates, solely based upon linear optics.

{\it Acknowledgements.}---
This work was supported by JST PRESTO Grant No.~JPMJPR23FA, JST Moonshot R\&D Grant No.~JPMJMS2064, JST Moonshot R\&D Grant No.~JPMJMS2061, UTokyo Foundation, and donations from Nichia Corporation. 
P.v.L. acknowledges financial support from the BMBF in Germany via QR.X, PhotonQ, QuKuK, and QuaPhySI, and from the EU’s Horizon Research and Innovation Actions (CLUSTEC).

\bibliography{ref.bib}

\begin{thebibliography}{76}%
\makeatletter
\providecommand \@ifxundefined [1]{%
 \@ifx{#1\undefined}
}%
\providecommand \@ifnum [1]{%
 \ifnum #1\expandafter \@firstoftwo
 \else \expandafter \@secondoftwo
 \fi
}%
\providecommand \@ifx [1]{%
 \ifx #1\expandafter \@firstoftwo
 \else \expandafter \@secondoftwo
 \fi
}%
\providecommand \natexlab [1]{#1}%
\providecommand \enquote  [1]{``#1''}%
\providecommand \bibnamefont  [1]{#1}%
\providecommand \bibfnamefont [1]{#1}%
\providecommand \citenamefont [1]{#1}%
\providecommand \href@noop [0]{\@secondoftwo}%
\providecommand \href [0]{\begingroup \@sanitize@url \@href}%
\providecommand \@href[1]{\@@startlink{#1}\@@href}%
\providecommand \@@href[1]{\endgroup#1\@@endlink}%
\providecommand \@sanitize@url [0]{\catcode `\\12\catcode `\$12\catcode
  `\&12\catcode `\#12\catcode `\^12\catcode `\_12\catcode `\%12\relax}%
\providecommand \@@startlink[1]{}%
\providecommand \@@endlink[0]{}%
\providecommand \url  [0]{\begingroup\@sanitize@url \@url }%
\providecommand \@url [1]{\endgroup\@href {#1}{\urlprefix }}%
\providecommand \urlprefix  [0]{URL }%
\providecommand \Eprint [0]{\href }%
\providecommand \doibase [0]{https://doi.org/}%
\providecommand \selectlanguage [0]{\@gobble}%
\providecommand \bibinfo  [0]{\@secondoftwo}%
\providecommand \bibfield  [0]{\@secondoftwo}%
\providecommand \translation [1]{[#1]}%
\providecommand \BibitemOpen [0]{}%
\providecommand \bibitemStop [0]{}%
\providecommand \bibitemNoStop [0]{.\EOS\space}%
\providecommand \EOS [0]{\spacefactor3000\relax}%
\providecommand \BibitemShut  [1]{\csname bibitem#1\endcsname}%
\let\auto@bib@innerbib\@empty
\bibitem [{\citenamefont {Shor}(1999)}]{shor1999polynomial}%
  \BibitemOpen
  \bibfield  {author} {\bibinfo {author} {\bibfnamefont {P.~W.}\ \bibnamefont
  {Shor}},\ }\bibfield  {title} {\bibinfo {title} {Polynomial-time algorithms
  for prime factorization and discrete logarithms on a quantum computer},\
  }\href@noop {} {\bibfield  {journal} {\bibinfo  {journal} {SIAM review}\
  }\textbf {\bibinfo {volume} {41}},\ \bibinfo {pages} {303} (\bibinfo {year}
  {1999})}\BibitemShut {NoStop}%
\bibitem [{\citenamefont {Grover}(1997)}]{grover1997quantum}%
  \BibitemOpen
  \bibfield  {author} {\bibinfo {author} {\bibfnamefont {L.~K.}\ \bibnamefont
  {Grover}},\ }\bibfield  {title} {\bibinfo {title} {Quantum mechanics helps in
  searching for a needle in a haystack},\ }\href@noop {} {\bibfield  {journal}
  {\bibinfo  {journal} {Physical review letters}\ }\textbf {\bibinfo {volume}
  {79}},\ \bibinfo {pages} {325} (\bibinfo {year} {1997})}\BibitemShut
  {NoStop}%
\bibitem [{\citenamefont {Ralph}\ \emph {et~al.}(2003)\citenamefont {Ralph},
  \citenamefont {Gilchrist}, \citenamefont {Milburn}, \citenamefont {Munro},\
  and\ \citenamefont {Glancy}}]{ralph2003quantum}%
  \BibitemOpen
  \bibfield  {author} {\bibinfo {author} {\bibfnamefont {T.~C.}\ \bibnamefont
  {Ralph}}, \bibinfo {author} {\bibfnamefont {A.}~\bibnamefont {Gilchrist}},
  \bibinfo {author} {\bibfnamefont {G.~J.}\ \bibnamefont {Milburn}}, \bibinfo
  {author} {\bibfnamefont {W.~J.}\ \bibnamefont {Munro}},\ and\ \bibinfo
  {author} {\bibfnamefont {S.}~\bibnamefont {Glancy}},\ }\bibfield  {title}
  {\bibinfo {title} {Quantum computation with optical coherent states},\
  }\href@noop {} {\bibfield  {journal} {\bibinfo  {journal} {Physical Review
  A}\ }\textbf {\bibinfo {volume} {68}},\ \bibinfo {pages} {042319} (\bibinfo
  {year} {2003})}\BibitemShut {NoStop}%
\bibitem [{\citenamefont {Michael}\ \emph {et~al.}(2016)\citenamefont
  {Michael}, \citenamefont {Silveri}, \citenamefont {Brierley}, \citenamefont
  {Albert}, \citenamefont {Salmilehto}, \citenamefont {Jiang},\ and\
  \citenamefont {Girvin}}]{michael2016new}%
  \BibitemOpen
  \bibfield  {author} {\bibinfo {author} {\bibfnamefont {M.~H.}\ \bibnamefont
  {Michael}}, \bibinfo {author} {\bibfnamefont {M.}~\bibnamefont {Silveri}},
  \bibinfo {author} {\bibfnamefont {R.}~\bibnamefont {Brierley}}, \bibinfo
  {author} {\bibfnamefont {V.~V.}\ \bibnamefont {Albert}}, \bibinfo {author}
  {\bibfnamefont {J.}~\bibnamefont {Salmilehto}}, \bibinfo {author}
  {\bibfnamefont {L.}~\bibnamefont {Jiang}},\ and\ \bibinfo {author}
  {\bibfnamefont {S.~M.}\ \bibnamefont {Girvin}},\ }\bibfield  {title}
  {\bibinfo {title} {New class of quantum error-correcting codes for a bosonic
  mode},\ }\href@noop {} {\bibfield  {journal} {\bibinfo  {journal} {Physical
  Review X}\ }\textbf {\bibinfo {volume} {6}},\ \bibinfo {pages} {031006}
  (\bibinfo {year} {2016})}\BibitemShut {NoStop}%
\bibitem [{\citenamefont {Gottesman}\ \emph {et~al.}(2001)\citenamefont
  {Gottesman}, \citenamefont {Kitaev},\ and\ \citenamefont
  {Preskill}}]{gottesman2001encoding}%
  \BibitemOpen
  \bibfield  {author} {\bibinfo {author} {\bibfnamefont {D.}~\bibnamefont
  {Gottesman}}, \bibinfo {author} {\bibfnamefont {A.}~\bibnamefont {Kitaev}},\
  and\ \bibinfo {author} {\bibfnamefont {J.}~\bibnamefont {Preskill}},\
  }\bibfield  {title} {\bibinfo {title} {Encoding a qubit in an oscillator},\
  }\href@noop {} {\bibfield  {journal} {\bibinfo  {journal} {Physical Review
  A}\ }\textbf {\bibinfo {volume} {64}},\ \bibinfo {pages} {012310} (\bibinfo
  {year} {2001})}\BibitemShut {NoStop}%
\bibitem [{\citenamefont {Li}\ \emph {et~al.}(2015)\citenamefont {Li},
  \citenamefont {Humphreys}, \citenamefont {Mendoza},\ and\ \citenamefont
  {Benjamin}}]{li2015resource}%
  \BibitemOpen
  \bibfield  {author} {\bibinfo {author} {\bibfnamefont {Y.}~\bibnamefont
  {Li}}, \bibinfo {author} {\bibfnamefont {P.~C.}\ \bibnamefont {Humphreys}},
  \bibinfo {author} {\bibfnamefont {G.~J.}\ \bibnamefont {Mendoza}},\ and\
  \bibinfo {author} {\bibfnamefont {S.~C.}\ \bibnamefont {Benjamin}},\
  }\bibfield  {title} {\bibinfo {title} {Resource costs for fault-tolerant
  linear optical quantum computing},\ }\href@noop {} {\bibfield  {journal}
  {\bibinfo  {journal} {Physical Review X}\ }\textbf {\bibinfo {volume} {5}},\
  \bibinfo {pages} {041007} (\bibinfo {year} {2015})}\BibitemShut {NoStop}%
\bibitem [{\citenamefont {Gimeno-Segovia}\ \emph {et~al.}(2015)\citenamefont
  {Gimeno-Segovia}, \citenamefont {Shadbolt}, \citenamefont {Browne},\ and\
  \citenamefont {Rudolph}}]{gimeno2015three}%
  \BibitemOpen
  \bibfield  {author} {\bibinfo {author} {\bibfnamefont {M.}~\bibnamefont
  {Gimeno-Segovia}}, \bibinfo {author} {\bibfnamefont {P.}~\bibnamefont
  {Shadbolt}}, \bibinfo {author} {\bibfnamefont {D.~E.}\ \bibnamefont
  {Browne}},\ and\ \bibinfo {author} {\bibfnamefont {T.}~\bibnamefont
  {Rudolph}},\ }\bibfield  {title} {\bibinfo {title} {From three-photon
  greenberger-horne-zeilinger states to ballistic universal quantum
  computation},\ }\href@noop {} {\bibfield  {journal} {\bibinfo  {journal}
  {Physical review letters}\ }\textbf {\bibinfo {volume} {115}},\ \bibinfo
  {pages} {020502} (\bibinfo {year} {2015})}\BibitemShut {NoStop}%
\bibitem [{\citenamefont {Bartolucci}\ \emph {et~al.}(2023)\citenamefont
  {Bartolucci}, \citenamefont {Birchall}, \citenamefont {Bombin}, \citenamefont
  {Cable}, \citenamefont {Dawson}, \citenamefont {Gimeno-Segovia},
  \citenamefont {Johnston}, \citenamefont {Kieling}, \citenamefont {Nickerson},
  \citenamefont {Pant} \emph {et~al.}}]{bartolucci2023fusion}%
  \BibitemOpen
  \bibfield  {author} {\bibinfo {author} {\bibfnamefont {S.}~\bibnamefont
  {Bartolucci}}, \bibinfo {author} {\bibfnamefont {P.}~\bibnamefont
  {Birchall}}, \bibinfo {author} {\bibfnamefont {H.}~\bibnamefont {Bombin}},
  \bibinfo {author} {\bibfnamefont {H.}~\bibnamefont {Cable}}, \bibinfo
  {author} {\bibfnamefont {C.}~\bibnamefont {Dawson}}, \bibinfo {author}
  {\bibfnamefont {M.}~\bibnamefont {Gimeno-Segovia}}, \bibinfo {author}
  {\bibfnamefont {E.}~\bibnamefont {Johnston}}, \bibinfo {author}
  {\bibfnamefont {K.}~\bibnamefont {Kieling}}, \bibinfo {author} {\bibfnamefont
  {N.}~\bibnamefont {Nickerson}}, \bibinfo {author} {\bibfnamefont
  {M.}~\bibnamefont {Pant}}, \emph {et~al.},\ }\bibfield  {title} {\bibinfo
  {title} {Fusion-based quantum computation},\ }\href@noop {} {\bibfield
  {journal} {\bibinfo  {journal} {Nature Communications}\ }\textbf {\bibinfo
  {volume} {14}},\ \bibinfo {pages} {912} (\bibinfo {year} {2023})}\BibitemShut
  {NoStop}%
\bibitem [{\citenamefont {Pankovich}\ \emph {et~al.}(2024)\citenamefont
  {Pankovich}, \citenamefont {Kan}, \citenamefont {Wan}, \citenamefont
  {Ostmann}, \citenamefont {Neville}, \citenamefont {Omkar}, \citenamefont
  {Sohbi},\ and\ \citenamefont {Br{\'a}dler}}]{pankovich2024high}%
  \BibitemOpen
  \bibfield  {author} {\bibinfo {author} {\bibfnamefont {B.}~\bibnamefont
  {Pankovich}}, \bibinfo {author} {\bibfnamefont {A.}~\bibnamefont {Kan}},
  \bibinfo {author} {\bibfnamefont {K.~H.}\ \bibnamefont {Wan}}, \bibinfo
  {author} {\bibfnamefont {M.}~\bibnamefont {Ostmann}}, \bibinfo {author}
  {\bibfnamefont {A.}~\bibnamefont {Neville}}, \bibinfo {author} {\bibfnamefont
  {S.}~\bibnamefont {Omkar}}, \bibinfo {author} {\bibfnamefont
  {A.}~\bibnamefont {Sohbi}},\ and\ \bibinfo {author} {\bibfnamefont
  {K.}~\bibnamefont {Br{\'a}dler}},\ }\bibfield  {title} {\bibinfo {title}
  {High-photon-loss threshold quantum computing using ghz-state measurements},\
  }\href@noop {} {\bibfield  {journal} {\bibinfo  {journal} {Physical Review
  Letters}\ }\textbf {\bibinfo {volume} {133}},\ \bibinfo {pages} {050604}
  (\bibinfo {year} {2024})}\BibitemShut {NoStop}%
\bibitem [{\citenamefont {Song}\ \emph {et~al.}(2024)\citenamefont {Song},
  \citenamefont {Kang}, \citenamefont {Kim},\ and\ \citenamefont
  {Lee}}]{song2024encoded}%
  \BibitemOpen
  \bibfield  {author} {\bibinfo {author} {\bibfnamefont {W.}~\bibnamefont
  {Song}}, \bibinfo {author} {\bibfnamefont {N.}~\bibnamefont {Kang}}, \bibinfo
  {author} {\bibfnamefont {Y.-S.}\ \bibnamefont {Kim}},\ and\ \bibinfo {author}
  {\bibfnamefont {S.-W.}\ \bibnamefont {Lee}},\ }\bibfield  {title} {\bibinfo
  {title} {Encoded-fusion-based quantum computation for high thresholds with
  linear optics},\ }\href@noop {} {\bibfield  {journal} {\bibinfo  {journal}
  {Physical Review Letters}\ }\textbf {\bibinfo {volume} {133}},\ \bibinfo
  {pages} {050605} (\bibinfo {year} {2024})}\BibitemShut {NoStop}%
\bibitem [{\citenamefont {Myers}\ and\ \citenamefont
  {Ralph}(2011)}]{myers2011coherent}%
  \BibitemOpen
  \bibfield  {author} {\bibinfo {author} {\bibfnamefont {C.~R.}\ \bibnamefont
  {Myers}}\ and\ \bibinfo {author} {\bibfnamefont {T.~C.}\ \bibnamefont
  {Ralph}},\ }\bibfield  {title} {\bibinfo {title} {Coherent state topological
  cluster state production},\ }\href@noop {} {\bibfield  {journal} {\bibinfo
  {journal} {New Journal of Physics}\ }\textbf {\bibinfo {volume} {13}},\
  \bibinfo {pages} {115015} (\bibinfo {year} {2011})}\BibitemShut {NoStop}%
\bibitem [{\citenamefont {Lund}\ \emph {et~al.}(2008)\citenamefont {Lund},
  \citenamefont {Ralph},\ and\ \citenamefont {Haselgrove}}]{lund2008fault}%
  \BibitemOpen
  \bibfield  {author} {\bibinfo {author} {\bibfnamefont {A.~P.}\ \bibnamefont
  {Lund}}, \bibinfo {author} {\bibfnamefont {T.~C.}\ \bibnamefont {Ralph}},\
  and\ \bibinfo {author} {\bibfnamefont {H.~L.}\ \bibnamefont {Haselgrove}},\
  }\bibfield  {title} {\bibinfo {title} {Fault-tolerant linear optical quantum
  computing with small-amplitude coherent states},\ }\href@noop {} {\bibfield
  {journal} {\bibinfo  {journal} {Physical review letters}\ }\textbf {\bibinfo
  {volume} {100}},\ \bibinfo {pages} {030503} (\bibinfo {year}
  {2008})}\BibitemShut {NoStop}%
\bibitem [{\citenamefont {Pantaleoni}\ \emph {et~al.}(2020)\citenamefont
  {Pantaleoni}, \citenamefont {Baragiola},\ and\ \citenamefont
  {Menicucci}}]{pantaleoni2020modular}%
  \BibitemOpen
  \bibfield  {author} {\bibinfo {author} {\bibfnamefont {G.}~\bibnamefont
  {Pantaleoni}}, \bibinfo {author} {\bibfnamefont {B.~Q.}\ \bibnamefont
  {Baragiola}},\ and\ \bibinfo {author} {\bibfnamefont {N.~C.}\ \bibnamefont
  {Menicucci}},\ }\bibfield  {title} {\bibinfo {title} {Modular bosonic
  subsystem codes},\ }\href@noop {} {\bibfield  {journal} {\bibinfo  {journal}
  {Physical Review Letters}\ }\textbf {\bibinfo {volume} {125}},\ \bibinfo
  {pages} {040501} (\bibinfo {year} {2020})}\BibitemShut {NoStop}%
\bibitem [{\citenamefont {Walshe}\ \emph {et~al.}(2020)\citenamefont {Walshe},
  \citenamefont {Baragiola}, \citenamefont {Alexander},\ and\ \citenamefont
  {Menicucci}}]{walshe2020continuous}%
  \BibitemOpen
  \bibfield  {author} {\bibinfo {author} {\bibfnamefont {B.~W.}\ \bibnamefont
  {Walshe}}, \bibinfo {author} {\bibfnamefont {B.~Q.}\ \bibnamefont
  {Baragiola}}, \bibinfo {author} {\bibfnamefont {R.~N.}\ \bibnamefont
  {Alexander}},\ and\ \bibinfo {author} {\bibfnamefont {N.~C.}\ \bibnamefont
  {Menicucci}},\ }\bibfield  {title} {\bibinfo {title} {Continuous-variable
  gate teleportation and bosonic-code error correction},\ }\href@noop {}
  {\bibfield  {journal} {\bibinfo  {journal} {Physical Review A}\ }\textbf
  {\bibinfo {volume} {102}},\ \bibinfo {pages} {062411} (\bibinfo {year}
  {2020})}\BibitemShut {NoStop}%
\bibitem [{\citenamefont {Bourassa}\ \emph {et~al.}(2021)\citenamefont
  {Bourassa}, \citenamefont {Alexander}, \citenamefont {Vasmer}, \citenamefont
  {Patil}, \citenamefont {Tzitrin}, \citenamefont {Matsuura}, \citenamefont
  {Su}, \citenamefont {Baragiola}, \citenamefont {Guha}, \citenamefont
  {Dauphinais} \emph {et~al.}}]{bourassa2021blueprint}%
  \BibitemOpen
  \bibfield  {author} {\bibinfo {author} {\bibfnamefont {J.~E.}\ \bibnamefont
  {Bourassa}}, \bibinfo {author} {\bibfnamefont {R.~N.}\ \bibnamefont
  {Alexander}}, \bibinfo {author} {\bibfnamefont {M.}~\bibnamefont {Vasmer}},
  \bibinfo {author} {\bibfnamefont {A.}~\bibnamefont {Patil}}, \bibinfo
  {author} {\bibfnamefont {I.}~\bibnamefont {Tzitrin}}, \bibinfo {author}
  {\bibfnamefont {T.}~\bibnamefont {Matsuura}}, \bibinfo {author}
  {\bibfnamefont {D.}~\bibnamefont {Su}}, \bibinfo {author} {\bibfnamefont
  {B.~Q.}\ \bibnamefont {Baragiola}}, \bibinfo {author} {\bibfnamefont
  {S.}~\bibnamefont {Guha}}, \bibinfo {author} {\bibfnamefont {G.}~\bibnamefont
  {Dauphinais}}, \emph {et~al.},\ }\bibfield  {title} {\bibinfo {title}
  {Blueprint for a scalable photonic fault-tolerant quantum computer},\
  }\href@noop {} {\bibfield  {journal} {\bibinfo  {journal} {Quantum}\ }\textbf
  {\bibinfo {volume} {5}},\ \bibinfo {pages} {392} (\bibinfo {year}
  {2021})}\BibitemShut {NoStop}%
\bibitem [{\citenamefont {Tzitrin}\ \emph {et~al.}(2021)\citenamefont
  {Tzitrin}, \citenamefont {Matsuura}, \citenamefont {Alexander}, \citenamefont
  {Dauphinais}, \citenamefont {Bourassa}, \citenamefont {Sabapathy},
  \citenamefont {Menicucci},\ and\ \citenamefont {Dhand}}]{tzitrin2021fault}%
  \BibitemOpen
  \bibfield  {author} {\bibinfo {author} {\bibfnamefont {I.}~\bibnamefont
  {Tzitrin}}, \bibinfo {author} {\bibfnamefont {T.}~\bibnamefont {Matsuura}},
  \bibinfo {author} {\bibfnamefont {R.~N.}\ \bibnamefont {Alexander}}, \bibinfo
  {author} {\bibfnamefont {G.}~\bibnamefont {Dauphinais}}, \bibinfo {author}
  {\bibfnamefont {J.~E.}\ \bibnamefont {Bourassa}}, \bibinfo {author}
  {\bibfnamefont {K.~K.}\ \bibnamefont {Sabapathy}}, \bibinfo {author}
  {\bibfnamefont {N.~C.}\ \bibnamefont {Menicucci}},\ and\ \bibinfo {author}
  {\bibfnamefont {I.}~\bibnamefont {Dhand}},\ }\bibfield  {title} {\bibinfo
  {title} {Fault-tolerant quantum computation with static linear optics},\
  }\href@noop {} {\bibfield  {journal} {\bibinfo  {journal} {PRX Quantum}\
  }\textbf {\bibinfo {volume} {2}},\ \bibinfo {pages} {040353} (\bibinfo {year}
  {2021})}\BibitemShut {NoStop}%
\bibitem [{\citenamefont {Larsen}\ \emph {et~al.}(2021)\citenamefont {Larsen},
  \citenamefont {Chamberland}, \citenamefont {Noh}, \citenamefont
  {Neergaard-Nielsen},\ and\ \citenamefont {Andersen}}]{larsen2021fault}%
  \BibitemOpen
  \bibfield  {author} {\bibinfo {author} {\bibfnamefont {M.~V.}\ \bibnamefont
  {Larsen}}, \bibinfo {author} {\bibfnamefont {C.}~\bibnamefont {Chamberland}},
  \bibinfo {author} {\bibfnamefont {K.}~\bibnamefont {Noh}}, \bibinfo {author}
  {\bibfnamefont {J.~S.}\ \bibnamefont {Neergaard-Nielsen}},\ and\ \bibinfo
  {author} {\bibfnamefont {U.~L.}\ \bibnamefont {Andersen}},\ }\bibfield
  {title} {\bibinfo {title} {Fault-tolerant continuous-variable
  measurement-based quantum computation architecture},\ }\href@noop {}
  {\bibfield  {journal} {\bibinfo  {journal} {PRX Quantum}\ }\textbf {\bibinfo
  {volume} {2}},\ \bibinfo {pages} {030325} (\bibinfo {year}
  {2021})}\BibitemShut {NoStop}%
\bibitem [{\citenamefont {Pantaleoni}\ \emph {et~al.}(2021)\citenamefont
  {Pantaleoni}, \citenamefont {Baragiola},\ and\ \citenamefont
  {Menicucci}}]{pantaleoni2021subsystem}%
  \BibitemOpen
  \bibfield  {author} {\bibinfo {author} {\bibfnamefont {G.}~\bibnamefont
  {Pantaleoni}}, \bibinfo {author} {\bibfnamefont {B.~Q.}\ \bibnamefont
  {Baragiola}},\ and\ \bibinfo {author} {\bibfnamefont {N.~C.}\ \bibnamefont
  {Menicucci}},\ }\bibfield  {title} {\bibinfo {title} {Subsystem analysis of
  continuous-variable resource states},\ }\href@noop {} {\bibfield  {journal}
  {\bibinfo  {journal} {Physical Review A}\ }\textbf {\bibinfo {volume}
  {104}},\ \bibinfo {pages} {012430} (\bibinfo {year} {2021})}\BibitemShut
  {NoStop}%
\bibitem [{\citenamefont {Fukui}\ and\ \citenamefont
  {Takeda}(2022)}]{fukui2022building}%
  \BibitemOpen
  \bibfield  {author} {\bibinfo {author} {\bibfnamefont {K.}~\bibnamefont
  {Fukui}}\ and\ \bibinfo {author} {\bibfnamefont {S.}~\bibnamefont {Takeda}},\
  }\bibfield  {title} {\bibinfo {title} {Building a large-scale quantum
  computer with continuous-variable optical technologies},\ }\href@noop {}
  {\bibfield  {journal} {\bibinfo  {journal} {Journal of Physics B: Atomic,
  Molecular and Optical Physics}\ }\textbf {\bibinfo {volume} {55}},\ \bibinfo
  {pages} {012001} (\bibinfo {year} {2022})}\BibitemShut {NoStop}%
\bibitem [{\citenamefont {Schmidt}\ and\ \citenamefont {van
  Loock}(2022)}]{schmidt2022quantum}%
  \BibitemOpen
  \bibfield  {author} {\bibinfo {author} {\bibfnamefont {F.}~\bibnamefont
  {Schmidt}}\ and\ \bibinfo {author} {\bibfnamefont {P.}~\bibnamefont {van
  Loock}},\ }\bibfield  {title} {\bibinfo {title} {Quantum error correction
  with higher gottesman-kitaev-preskill codes: Minimal measurements and linear
  optics},\ }\href@noop {} {\bibfield  {journal} {\bibinfo  {journal} {Physical
  Review A}\ }\textbf {\bibinfo {volume} {105}},\ \bibinfo {pages} {042427}
  (\bibinfo {year} {2022})}\BibitemShut {NoStop}%
\bibitem [{\citenamefont {Asavanant}\ \emph {et~al.}(2023)\citenamefont
  {Asavanant}, \citenamefont {Fukui}, \citenamefont {Sakaguchi},\ and\
  \citenamefont {Furusawa}}]{asavanant2023switching}%
  \BibitemOpen
  \bibfield  {author} {\bibinfo {author} {\bibfnamefont {W.}~\bibnamefont
  {Asavanant}}, \bibinfo {author} {\bibfnamefont {K.}~\bibnamefont {Fukui}},
  \bibinfo {author} {\bibfnamefont {A.}~\bibnamefont {Sakaguchi}},\ and\
  \bibinfo {author} {\bibfnamefont {A.}~\bibnamefont {Furusawa}},\ }\bibfield
  {title} {\bibinfo {title} {Switching-free time-domain optical quantum
  computation with quantum teleportation},\ }\href@noop {} {\bibfield
  {journal} {\bibinfo  {journal} {Physical Review A}\ }\textbf {\bibinfo
  {volume} {107}},\ \bibinfo {pages} {032412} (\bibinfo {year}
  {2023})}\BibitemShut {NoStop}%
\bibitem [{\citenamefont {Stafford}\ and\ \citenamefont
  {Menicucci}(2023)}]{stafford2023biased}%
  \BibitemOpen
  \bibfield  {author} {\bibinfo {author} {\bibfnamefont {M.~P.}\ \bibnamefont
  {Stafford}}\ and\ \bibinfo {author} {\bibfnamefont {N.~C.}\ \bibnamefont
  {Menicucci}},\ }\bibfield  {title} {\bibinfo {title} {Biased
  gottesman-kitaev-preskill repetition code},\ }\href@noop {} {\bibfield
  {journal} {\bibinfo  {journal} {Physical Review A}\ }\textbf {\bibinfo
  {volume} {108}},\ \bibinfo {pages} {052428} (\bibinfo {year}
  {2023})}\BibitemShut {NoStop}%
\bibitem [{\citenamefont {Walshe}\ \emph {et~al.}(2024)\citenamefont {Walshe},
  \citenamefont {Baragiola}, \citenamefont {Ferretti}, \citenamefont {Gefaell},
  \citenamefont {Vasmer}, \citenamefont {Weil}, \citenamefont {Matsuura},
  \citenamefont {Jaeken}, \citenamefont {Pantaleoni}, \citenamefont {Han} \emph
  {et~al.}}]{walshe2024linear}%
  \BibitemOpen
  \bibfield  {author} {\bibinfo {author} {\bibfnamefont {B.~W.}\ \bibnamefont
  {Walshe}}, \bibinfo {author} {\bibfnamefont {B.~Q.}\ \bibnamefont
  {Baragiola}}, \bibinfo {author} {\bibfnamefont {H.}~\bibnamefont {Ferretti}},
  \bibinfo {author} {\bibfnamefont {J.}~\bibnamefont {Gefaell}}, \bibinfo
  {author} {\bibfnamefont {M.}~\bibnamefont {Vasmer}}, \bibinfo {author}
  {\bibfnamefont {R.}~\bibnamefont {Weil}}, \bibinfo {author} {\bibfnamefont
  {T.}~\bibnamefont {Matsuura}}, \bibinfo {author} {\bibfnamefont
  {T.}~\bibnamefont {Jaeken}}, \bibinfo {author} {\bibfnamefont
  {G.}~\bibnamefont {Pantaleoni}}, \bibinfo {author} {\bibfnamefont
  {Z.}~\bibnamefont {Han}}, \emph {et~al.},\ }\bibfield  {title} {\bibinfo
  {title} {Linear-optical quantum computation with arbitrary error-correcting
  codes},\ }\href@noop {} {\bibfield  {journal} {\bibinfo  {journal} {arXiv
  preprint arXiv:2408.04126}\ } (\bibinfo {year} {2024})}\BibitemShut {NoStop}%
\bibitem [{\citenamefont {Menicucci}(2014)}]{menicucci2014fault}%
  \BibitemOpen
  \bibfield  {author} {\bibinfo {author} {\bibfnamefont {N.~C.}\ \bibnamefont
  {Menicucci}},\ }\bibfield  {title} {\bibinfo {title} {Fault-tolerant
  measurement-based quantum computing with continuous-variable cluster
  states},\ }\href@noop {} {\bibfield  {journal} {\bibinfo  {journal} {Physical
  review letters}\ }\textbf {\bibinfo {volume} {112}},\ \bibinfo {pages}
  {120504} (\bibinfo {year} {2014})}\BibitemShut {NoStop}%
\bibitem [{\citenamefont {Albert}\ \emph {et~al.}(2018)\citenamefont {Albert},
  \citenamefont {Noh}, \citenamefont {Duivenvoorden}, \citenamefont {Young},
  \citenamefont {Brierley}, \citenamefont {Reinhold}, \citenamefont {Vuillot},
  \citenamefont {Li}, \citenamefont {Shen}, \citenamefont {Girvin} \emph
  {et~al.}}]{albert2018performance}%
  \BibitemOpen
  \bibfield  {author} {\bibinfo {author} {\bibfnamefont {V.~V.}\ \bibnamefont
  {Albert}}, \bibinfo {author} {\bibfnamefont {K.}~\bibnamefont {Noh}},
  \bibinfo {author} {\bibfnamefont {K.}~\bibnamefont {Duivenvoorden}}, \bibinfo
  {author} {\bibfnamefont {D.~J.}\ \bibnamefont {Young}}, \bibinfo {author}
  {\bibfnamefont {R.}~\bibnamefont {Brierley}}, \bibinfo {author}
  {\bibfnamefont {P.}~\bibnamefont {Reinhold}}, \bibinfo {author}
  {\bibfnamefont {C.}~\bibnamefont {Vuillot}}, \bibinfo {author} {\bibfnamefont
  {L.}~\bibnamefont {Li}}, \bibinfo {author} {\bibfnamefont {C.}~\bibnamefont
  {Shen}}, \bibinfo {author} {\bibfnamefont {S.}~\bibnamefont {Girvin}}, \emph
  {et~al.},\ }\bibfield  {title} {\bibinfo {title} {Performance and structure
  of single-mode bosonic codes},\ }\href@noop {} {\bibfield  {journal}
  {\bibinfo  {journal} {Physical Review A}\ }\textbf {\bibinfo {volume} {97}},\
  \bibinfo {pages} {032346} (\bibinfo {year} {2018})}\BibitemShut {NoStop}%
\bibitem [{\citenamefont {Baragiola}\ \emph {et~al.}(2019)\citenamefont
  {Baragiola}, \citenamefont {Pantaleoni}, \citenamefont {Alexander},
  \citenamefont {Karanjai},\ and\ \citenamefont {Menicucci}}]{19Baragiola}%
  \BibitemOpen
  \bibfield  {author} {\bibinfo {author} {\bibfnamefont {B.~Q.}\ \bibnamefont
  {Baragiola}}, \bibinfo {author} {\bibfnamefont {G.}~\bibnamefont
  {Pantaleoni}}, \bibinfo {author} {\bibfnamefont {R.~N.}\ \bibnamefont
  {Alexander}}, \bibinfo {author} {\bibfnamefont {A.}~\bibnamefont
  {Karanjai}},\ and\ \bibinfo {author} {\bibfnamefont {N.~C.}\ \bibnamefont
  {Menicucci}},\ }\bibfield  {title} {\bibinfo {title} {All-gaussian
  universality and fault tolerance with the gottesman-kitaev-preskill code},\
  }\href {https://doi.org/10.1103/PhysRevLett.123.200502} {\bibfield  {journal}
  {\bibinfo  {journal} {Phys. Rev. Lett.}\ }\textbf {\bibinfo {volume} {123}},\
  \bibinfo {pages} {200502} (\bibinfo {year} {2019})}\BibitemShut {NoStop}%
\bibitem [{\citenamefont {Fukui}\ \emph {et~al.}(2017)\citenamefont {Fukui},
  \citenamefont {Tomita},\ and\ \citenamefont {Okamoto}}]{fukui2017analog}%
  \BibitemOpen
  \bibfield  {author} {\bibinfo {author} {\bibfnamefont {K.}~\bibnamefont
  {Fukui}}, \bibinfo {author} {\bibfnamefont {A.}~\bibnamefont {Tomita}},\ and\
  \bibinfo {author} {\bibfnamefont {A.}~\bibnamefont {Okamoto}},\ }\bibfield
  {title} {\bibinfo {title} {Analog quantum error correction with encoding a
  qubit into an oscillator},\ }\href@noop {} {\bibfield  {journal} {\bibinfo
  {journal} {Physical review letters}\ }\textbf {\bibinfo {volume} {119}},\
  \bibinfo {pages} {180507} (\bibinfo {year} {2017})}\BibitemShut {NoStop}%
\bibitem [{\citenamefont {Fukui}\ \emph {et~al.}(2018)\citenamefont {Fukui},
  \citenamefont {Tomita}, \citenamefont {Okamoto},\ and\ \citenamefont
  {Fujii}}]{fukui2018high}%
  \BibitemOpen
  \bibfield  {author} {\bibinfo {author} {\bibfnamefont {K.}~\bibnamefont
  {Fukui}}, \bibinfo {author} {\bibfnamefont {A.}~\bibnamefont {Tomita}},
  \bibinfo {author} {\bibfnamefont {A.}~\bibnamefont {Okamoto}},\ and\ \bibinfo
  {author} {\bibfnamefont {K.}~\bibnamefont {Fujii}},\ }\bibfield  {title}
  {\bibinfo {title} {High-threshold fault-tolerant quantum computation with
  analog quantum error correction},\ }\href@noop {} {\bibfield  {journal}
  {\bibinfo  {journal} {Physical review X}\ }\textbf {\bibinfo {volume} {8}},\
  \bibinfo {pages} {021054} (\bibinfo {year} {2018})}\BibitemShut {NoStop}%
\bibitem [{\citenamefont {Konno}\ \emph {et~al.}(2024)\citenamefont {Konno},
  \citenamefont {Asavanant}, \citenamefont {Hanamura}, \citenamefont
  {Nagayoshi}, \citenamefont {Fukui}, \citenamefont {Sakaguchi}, \citenamefont
  {Ide}, \citenamefont {China}, \citenamefont {Yabuno}, \citenamefont {Miki}
  \emph {et~al.}}]{konno2024logical}%
  \BibitemOpen
  \bibfield  {author} {\bibinfo {author} {\bibfnamefont {S.}~\bibnamefont
  {Konno}}, \bibinfo {author} {\bibfnamefont {W.}~\bibnamefont {Asavanant}},
  \bibinfo {author} {\bibfnamefont {F.}~\bibnamefont {Hanamura}}, \bibinfo
  {author} {\bibfnamefont {H.}~\bibnamefont {Nagayoshi}}, \bibinfo {author}
  {\bibfnamefont {K.}~\bibnamefont {Fukui}}, \bibinfo {author} {\bibfnamefont
  {A.}~\bibnamefont {Sakaguchi}}, \bibinfo {author} {\bibfnamefont
  {R.}~\bibnamefont {Ide}}, \bibinfo {author} {\bibfnamefont {F.}~\bibnamefont
  {China}}, \bibinfo {author} {\bibfnamefont {M.}~\bibnamefont {Yabuno}},
  \bibinfo {author} {\bibfnamefont {S.}~\bibnamefont {Miki}}, \emph {et~al.},\
  }\bibfield  {title} {\bibinfo {title} {Logical states for fault-tolerant
  quantum computation with propagating light},\ }\href@noop {} {\bibfield
  {journal} {\bibinfo  {journal} {Science}\ }\textbf {\bibinfo {volume}
  {383}},\ \bibinfo {pages} {289} (\bibinfo {year} {2024})}\BibitemShut
  {NoStop}%
\bibitem [{\citenamefont {van Loock}(2011)}]{van2011optical}%
  \BibitemOpen
  \bibfield  {author} {\bibinfo {author} {\bibfnamefont {P.}~\bibnamefont {van
  Loock}},\ }\bibfield  {title} {\bibinfo {title} {Optical hybrid approaches to
  quantum information},\ }\href@noop {} {\bibfield  {journal} {\bibinfo
  {journal} {Laser \& Photonics Reviews}\ }\textbf {\bibinfo {volume} {5}},\
  \bibinfo {pages} {167} (\bibinfo {year} {2011})}\BibitemShut {NoStop}%
\bibitem [{\citenamefont {Andersen}\ \emph {et~al.}(2015)\citenamefont
  {Andersen}, \citenamefont {Neergaard-Nielsen}, \citenamefont {van Loock},\
  and\ \citenamefont {Furusawa}}]{andersen2015hybrid}%
  \BibitemOpen
  \bibfield  {author} {\bibinfo {author} {\bibfnamefont {U.~L.}\ \bibnamefont
  {Andersen}}, \bibinfo {author} {\bibfnamefont {J.~S.}\ \bibnamefont
  {Neergaard-Nielsen}}, \bibinfo {author} {\bibfnamefont {P.}~\bibnamefont {van
  Loock}},\ and\ \bibinfo {author} {\bibfnamefont {A.}~\bibnamefont
  {Furusawa}},\ }\bibfield  {title} {\bibinfo {title} {Hybrid discrete-and
  continuous-variable quantum information},\ }\href@noop {} {\bibfield
  {journal} {\bibinfo  {journal} {Nature Physics}\ }\textbf {\bibinfo {volume}
  {11}},\ \bibinfo {pages} {713} (\bibinfo {year} {2015})}\BibitemShut
  {NoStop}%
\bibitem [{\citenamefont {Lee}\ and\ \citenamefont
  {Jeong}(2013)}]{lee2013near}%
  \BibitemOpen
  \bibfield  {author} {\bibinfo {author} {\bibfnamefont {S.-W.}\ \bibnamefont
  {Lee}}\ and\ \bibinfo {author} {\bibfnamefont {H.}~\bibnamefont {Jeong}},\
  }\bibfield  {title} {\bibinfo {title} {Near-deterministic quantum
  teleportation and resource-efficient quantum computation using linear optics
  and hybrid qubits},\ }\href@noop {} {\bibfield  {journal} {\bibinfo
  {journal} {Physical Review A}\ }\textbf {\bibinfo {volume} {87}},\ \bibinfo
  {pages} {022326} (\bibinfo {year} {2013})}\BibitemShut {NoStop}%
\bibitem [{\citenamefont {Jeong}\ \emph {et~al.}(2014)\citenamefont {Jeong},
  \citenamefont {Zavatta}, \citenamefont {Kang}, \citenamefont {Lee},
  \citenamefont {Costanzo}, \citenamefont {Grandi}, \citenamefont {Ralph},\
  and\ \citenamefont {Bellini}}]{jeong2014generation}%
  \BibitemOpen
  \bibfield  {author} {\bibinfo {author} {\bibfnamefont {H.}~\bibnamefont
  {Jeong}}, \bibinfo {author} {\bibfnamefont {A.}~\bibnamefont {Zavatta}},
  \bibinfo {author} {\bibfnamefont {M.}~\bibnamefont {Kang}}, \bibinfo {author}
  {\bibfnamefont {S.-W.}\ \bibnamefont {Lee}}, \bibinfo {author} {\bibfnamefont
  {L.~S.}\ \bibnamefont {Costanzo}}, \bibinfo {author} {\bibfnamefont
  {S.}~\bibnamefont {Grandi}}, \bibinfo {author} {\bibfnamefont {T.~C.}\
  \bibnamefont {Ralph}},\ and\ \bibinfo {author} {\bibfnamefont
  {M.}~\bibnamefont {Bellini}},\ }\bibfield  {title} {\bibinfo {title}
  {Generation of hybrid entanglement of light},\ }\href@noop {} {\bibfield
  {journal} {\bibinfo  {journal} {Nature Photonics}\ }\textbf {\bibinfo
  {volume} {8}},\ \bibinfo {pages} {564} (\bibinfo {year} {2014})}\BibitemShut
  {NoStop}%
\bibitem [{\citenamefont {Lee}\ \emph {et~al.}(2015)\citenamefont {Lee},
  \citenamefont {Park}, \citenamefont {Ralph},\ and\ \citenamefont
  {Jeong}}]{lee2015nearly}%
  \BibitemOpen
  \bibfield  {author} {\bibinfo {author} {\bibfnamefont {S.-W.}\ \bibnamefont
  {Lee}}, \bibinfo {author} {\bibfnamefont {K.}~\bibnamefont {Park}}, \bibinfo
  {author} {\bibfnamefont {T.~C.}\ \bibnamefont {Ralph}},\ and\ \bibinfo
  {author} {\bibfnamefont {H.}~\bibnamefont {Jeong}},\ }\bibfield  {title}
  {\bibinfo {title} {Nearly deterministic bell measurement for multiphoton
  qubits and its application to quantum information processing},\ }\href@noop
  {} {\bibfield  {journal} {\bibinfo  {journal} {Physical review letters}\
  }\textbf {\bibinfo {volume} {114}},\ \bibinfo {pages} {113603} (\bibinfo
  {year} {2015})}\BibitemShut {NoStop}%
\bibitem [{\citenamefont {Omkar}\ \emph {et~al.}(2020)\citenamefont {Omkar},
  \citenamefont {Teo},\ and\ \citenamefont {Jeong}}]{omkar2020resource}%
  \BibitemOpen
  \bibfield  {author} {\bibinfo {author} {\bibfnamefont {S.}~\bibnamefont
  {Omkar}}, \bibinfo {author} {\bibfnamefont {Y.~S.}\ \bibnamefont {Teo}},\
  and\ \bibinfo {author} {\bibfnamefont {H.}~\bibnamefont {Jeong}},\ }\bibfield
   {title} {\bibinfo {title} {Resource-efficient topological fault-tolerant
  quantum computation with hybrid entanglement of light},\ }\href@noop {}
  {\bibfield  {journal} {\bibinfo  {journal} {Physical Review Letters}\
  }\textbf {\bibinfo {volume} {125}},\ \bibinfo {pages} {060501} (\bibinfo
  {year} {2020})}\BibitemShut {NoStop}%
\bibitem [{\citenamefont {Omkar}\ \emph {et~al.}(2021)\citenamefont {Omkar},
  \citenamefont {Teo}, \citenamefont {Lee},\ and\ \citenamefont
  {Jeong}}]{omkar2021highly}%
  \BibitemOpen
  \bibfield  {author} {\bibinfo {author} {\bibfnamefont {S.}~\bibnamefont
  {Omkar}}, \bibinfo {author} {\bibfnamefont {Y.}~\bibnamefont {Teo}}, \bibinfo
  {author} {\bibfnamefont {S.-W.}\ \bibnamefont {Lee}},\ and\ \bibinfo {author}
  {\bibfnamefont {H.}~\bibnamefont {Jeong}},\ }\bibfield  {title} {\bibinfo
  {title} {Highly photon-loss-tolerant quantum computing using hybrid qubits},\
  }\href@noop {} {\bibfield  {journal} {\bibinfo  {journal} {Physical Review
  A}\ }\textbf {\bibinfo {volume} {103}},\ \bibinfo {pages} {032602} (\bibinfo
  {year} {2021})}\BibitemShut {NoStop}%
\bibitem [{\citenamefont {Lee}\ \emph {et~al.}(2024)\citenamefont {Lee},
  \citenamefont {Kang}, \citenamefont {Lee}, \citenamefont {Jeong},
  \citenamefont {Jiang},\ and\ \citenamefont {Lee}}]{lee2024fault}%
  \BibitemOpen
  \bibfield  {author} {\bibinfo {author} {\bibfnamefont {J.}~\bibnamefont
  {Lee}}, \bibinfo {author} {\bibfnamefont {N.}~\bibnamefont {Kang}}, \bibinfo
  {author} {\bibfnamefont {S.-H.}\ \bibnamefont {Lee}}, \bibinfo {author}
  {\bibfnamefont {H.}~\bibnamefont {Jeong}}, \bibinfo {author} {\bibfnamefont
  {L.}~\bibnamefont {Jiang}},\ and\ \bibinfo {author} {\bibfnamefont {S.-W.}\
  \bibnamefont {Lee}},\ }\bibfield  {title} {\bibinfo {title} {Fault-tolerant
  quantum computation by hybrid qubits with bosonic cat code and single
  photons},\ }\href@noop {} {\bibfield  {journal} {\bibinfo  {journal} {PRX
  Quantum}\ }\textbf {\bibinfo {volume} {5}},\ \bibinfo {pages} {030322}
  (\bibinfo {year} {2024})}\BibitemShut {NoStop}%
\bibitem [{\citenamefont {Nemoto}\ and\ \citenamefont
  {Munro}(2004)}]{nemoto2004nearly}%
  \BibitemOpen
  \bibfield  {author} {\bibinfo {author} {\bibfnamefont {K.}~\bibnamefont
  {Nemoto}}\ and\ \bibinfo {author} {\bibfnamefont {W.~J.}\ \bibnamefont
  {Munro}},\ }\bibfield  {title} {\bibinfo {title} {Nearly deterministic linear
  optical controlled-not gate},\ }\href@noop {} {\bibfield  {journal} {\bibinfo
   {journal} {Physical review letters}\ }\textbf {\bibinfo {volume} {93}},\
  \bibinfo {pages} {250502} (\bibinfo {year} {2004})}\BibitemShut {NoStop}%
\bibitem [{\citenamefont {Yamaguchi}\ \emph {et~al.}(2006)\citenamefont
  {Yamaguchi}, \citenamefont {Nemoto},\ and\ \citenamefont
  {Munro}}]{yamaguchi2006quantum}%
  \BibitemOpen
  \bibfield  {author} {\bibinfo {author} {\bibfnamefont {F.}~\bibnamefont
  {Yamaguchi}}, \bibinfo {author} {\bibfnamefont {K.}~\bibnamefont {Nemoto}},\
  and\ \bibinfo {author} {\bibfnamefont {W.~J.}\ \bibnamefont {Munro}},\
  }\bibfield  {title} {\bibinfo {title} {Quantum error correction via robust
  probe modes},\ }\href@noop {} {\bibfield  {journal} {\bibinfo  {journal}
  {Physical Review A—Atomic, Molecular, and Optical Physics}\ }\textbf
  {\bibinfo {volume} {73}},\ \bibinfo {pages} {060302} (\bibinfo {year}
  {2006})}\BibitemShut {NoStop}%
\bibitem [{\citenamefont {Myers}\ \emph {et~al.}(2007)\citenamefont {Myers},
  \citenamefont {Silva}, \citenamefont {Nemoto},\ and\ \citenamefont
  {Munro}}]{myers2007stabilizer}%
  \BibitemOpen
  \bibfield  {author} {\bibinfo {author} {\bibfnamefont {C.~R.}\ \bibnamefont
  {Myers}}, \bibinfo {author} {\bibfnamefont {M.}~\bibnamefont {Silva}},
  \bibinfo {author} {\bibfnamefont {K.}~\bibnamefont {Nemoto}},\ and\ \bibinfo
  {author} {\bibfnamefont {W.~J.}\ \bibnamefont {Munro}},\ }\bibfield  {title}
  {\bibinfo {title} {Stabilizer quantum error correction with quantum bus
  computation},\ }\href@noop {} {\bibfield  {journal} {\bibinfo  {journal}
  {Physical Review A—Atomic, Molecular, and Optical Physics}\ }\textbf
  {\bibinfo {volume} {76}},\ \bibinfo {pages} {012303} (\bibinfo {year}
  {2007})}\BibitemShut {NoStop}%
\bibitem [{\citenamefont {Munro}\ \emph {et~al.}(2005)\citenamefont {Munro},
  \citenamefont {Nemoto},\ and\ \citenamefont {Spiller}}]{munro2005weak}%
  \BibitemOpen
  \bibfield  {author} {\bibinfo {author} {\bibfnamefont {W.~J.}\ \bibnamefont
  {Munro}}, \bibinfo {author} {\bibfnamefont {K.}~\bibnamefont {Nemoto}},\ and\
  \bibinfo {author} {\bibfnamefont {T.~P.}\ \bibnamefont {Spiller}},\
  }\bibfield  {title} {\bibinfo {title} {Weak nonlinearities: a new route to
  optical quantum computation},\ }\href@noop {} {\bibfield  {journal} {\bibinfo
   {journal} {New Journal of Physics}\ }\textbf {\bibinfo {volume} {7}},\
  \bibinfo {pages} {137} (\bibinfo {year} {2005})}\BibitemShut {NoStop}%
\bibitem [{\citenamefont {Spiller}\ \emph {et~al.}(2006)\citenamefont
  {Spiller}, \citenamefont {Nemoto}, \citenamefont {Braunstein}, \citenamefont
  {Munro}, \citenamefont {van Loock},\ and\ \citenamefont
  {Milburn}}]{spiller2006quantum}%
  \BibitemOpen
  \bibfield  {author} {\bibinfo {author} {\bibfnamefont {T.~P.}\ \bibnamefont
  {Spiller}}, \bibinfo {author} {\bibfnamefont {K.}~\bibnamefont {Nemoto}},
  \bibinfo {author} {\bibfnamefont {S.~L.}\ \bibnamefont {Braunstein}},
  \bibinfo {author} {\bibfnamefont {W.~J.}\ \bibnamefont {Munro}}, \bibinfo
  {author} {\bibfnamefont {P.}~\bibnamefont {van Loock}},\ and\ \bibinfo
  {author} {\bibfnamefont {G.~J.}\ \bibnamefont {Milburn}},\ }\bibfield
  {title} {\bibinfo {title} {Quantum computation by communication},\
  }\href@noop {} {\bibfield  {journal} {\bibinfo  {journal} {New Journal of
  Physics}\ }\textbf {\bibinfo {volume} {8}},\ \bibinfo {pages} {30} (\bibinfo
  {year} {2006})}\BibitemShut {NoStop}%
\bibitem [{\citenamefont {van Loock}\ \emph {et~al.}(2008)\citenamefont {van
  Loock}, \citenamefont {Munro}, \citenamefont {Nemoto}, \citenamefont
  {Spiller}, \citenamefont {Ladd}, \citenamefont {Braunstein},\ and\
  \citenamefont {Milburn}}]{van2008hybrid}%
  \BibitemOpen
  \bibfield  {author} {\bibinfo {author} {\bibfnamefont {P.}~\bibnamefont {van
  Loock}}, \bibinfo {author} {\bibfnamefont {W.}~\bibnamefont {Munro}},
  \bibinfo {author} {\bibfnamefont {K.}~\bibnamefont {Nemoto}}, \bibinfo
  {author} {\bibfnamefont {T.}~\bibnamefont {Spiller}}, \bibinfo {author}
  {\bibfnamefont {T.}~\bibnamefont {Ladd}}, \bibinfo {author} {\bibfnamefont
  {S.~L.}\ \bibnamefont {Braunstein}},\ and\ \bibinfo {author} {\bibfnamefont
  {G.}~\bibnamefont {Milburn}},\ }\bibfield  {title} {\bibinfo {title} {Hybrid
  quantum computation in quantum optics},\ }\href@noop {} {\bibfield  {journal}
  {\bibinfo  {journal} {Physical Review A—Atomic, Molecular, and Optical
  Physics}\ }\textbf {\bibinfo {volume} {78}},\ \bibinfo {pages} {022303}
  (\bibinfo {year} {2008})}\BibitemShut {NoStop}%
\bibitem [{\citenamefont {Venkataraman}\ \emph {et~al.}(2013)\citenamefont
  {Venkataraman}, \citenamefont {Saha},\ and\ \citenamefont
  {Gaeta}}]{venkataraman2013phase}%
  \BibitemOpen
  \bibfield  {author} {\bibinfo {author} {\bibfnamefont {V.}~\bibnamefont
  {Venkataraman}}, \bibinfo {author} {\bibfnamefont {K.}~\bibnamefont {Saha}},\
  and\ \bibinfo {author} {\bibfnamefont {A.~L.}\ \bibnamefont {Gaeta}},\
  }\bibfield  {title} {\bibinfo {title} {Phase modulation at the few-photon
  level for weak-nonlinearity-based quantum computing},\ }\href@noop {}
  {\bibfield  {journal} {\bibinfo  {journal} {Nature Photonics}\ }\textbf
  {\bibinfo {volume} {7}},\ \bibinfo {pages} {138} (\bibinfo {year}
  {2013})}\BibitemShut {NoStop}%
\bibitem [{\citenamefont {Feizpour}\ \emph {et~al.}(2015)\citenamefont
  {Feizpour}, \citenamefont {Hallaji}, \citenamefont {Dmochowski},\ and\
  \citenamefont {Steinberg}}]{feizpour2015observation}%
  \BibitemOpen
  \bibfield  {author} {\bibinfo {author} {\bibfnamefont {A.}~\bibnamefont
  {Feizpour}}, \bibinfo {author} {\bibfnamefont {M.}~\bibnamefont {Hallaji}},
  \bibinfo {author} {\bibfnamefont {G.}~\bibnamefont {Dmochowski}},\ and\
  \bibinfo {author} {\bibfnamefont {A.~M.}\ \bibnamefont {Steinberg}},\
  }\bibfield  {title} {\bibinfo {title} {Observation of the nonlinear phase
  shift due to single post-selected photons},\ }\href@noop {} {\bibfield
  {journal} {\bibinfo  {journal} {Nature Physics}\ }\textbf {\bibinfo {volume}
  {11}},\ \bibinfo {pages} {905} (\bibinfo {year} {2015})}\BibitemShut
  {NoStop}%
\bibitem [{not({\natexlab{a}})}]{note3}%
  \BibitemOpen
  \href@noop {} {}\bibinfo {note} {{T}he main obstacle to improve the threshold
  comes from the continuous-variable noise propagation through
  continuous-variable gates, while continuous-variable gates are essential to
  prepare a large-scale cluster state for measurement-based quantum computation
  with continuous-variables.}\BibitemShut {Stop}%
\bibitem [{\citenamefont {Shapiro}(2006)}]{shapiro2006single}%
  \BibitemOpen
  \bibfield  {author} {\bibinfo {author} {\bibfnamefont {J.~H.}\ \bibnamefont
  {Shapiro}},\ }\bibfield  {title} {\bibinfo {title} {Single-photon kerr
  nonlinearities do not help quantum computation},\ }\href@noop {} {\bibfield
  {journal} {\bibinfo  {journal} {Physical Review A—Atomic, Molecular, and
  Optical Physics}\ }\textbf {\bibinfo {volume} {73}},\ \bibinfo {pages}
  {062305} (\bibinfo {year} {2006})}\BibitemShut {NoStop}%
\bibitem [{\citenamefont {Gea-Banacloche}(2010)}]{gea2010impossibility}%
  \BibitemOpen
  \bibfield  {author} {\bibinfo {author} {\bibfnamefont {J.}~\bibnamefont
  {Gea-Banacloche}},\ }\bibfield  {title} {\bibinfo {title} {Impossibility of
  large phase shifts via the giant kerr effect with single-photon wave
  packets},\ }\href@noop {} {\bibfield  {journal} {\bibinfo  {journal}
  {Physical Review A—Atomic, Molecular, and Optical Physics}\ }\textbf
  {\bibinfo {volume} {81}},\ \bibinfo {pages} {043823} (\bibinfo {year}
  {2010})}\BibitemShut {NoStop}%
\bibitem [{\citenamefont {Yanagimoto}\ \emph
  {et~al.}(2023{\natexlab{a}})\citenamefont {Yanagimoto}, \citenamefont {Ng},
  \citenamefont {Jankowski}, \citenamefont {Nehra}, \citenamefont {McKenna},
  \citenamefont {Onodera}, \citenamefont {Wright}, \citenamefont {Hamerly},
  \citenamefont {Marandi}, \citenamefont {Fejer} \emph
  {et~al.}}]{yanagimoto2023mesoscopic}%
  \BibitemOpen
  \bibfield  {author} {\bibinfo {author} {\bibfnamefont {R.}~\bibnamefont
  {Yanagimoto}}, \bibinfo {author} {\bibfnamefont {E.}~\bibnamefont {Ng}},
  \bibinfo {author} {\bibfnamefont {M.}~\bibnamefont {Jankowski}}, \bibinfo
  {author} {\bibfnamefont {R.}~\bibnamefont {Nehra}}, \bibinfo {author}
  {\bibfnamefont {T.~P.}\ \bibnamefont {McKenna}}, \bibinfo {author}
  {\bibfnamefont {T.}~\bibnamefont {Onodera}}, \bibinfo {author} {\bibfnamefont
  {L.~G.}\ \bibnamefont {Wright}}, \bibinfo {author} {\bibfnamefont
  {R.}~\bibnamefont {Hamerly}}, \bibinfo {author} {\bibfnamefont
  {A.}~\bibnamefont {Marandi}}, \bibinfo {author} {\bibfnamefont
  {M.}~\bibnamefont {Fejer}}, \emph {et~al.},\ }\bibfield  {title} {\bibinfo
  {title} {Mesoscopic ultrafast nonlinear optics--the emergence of multimode
  quantum non-gaussian physics},\ }\href@noop {} {\bibfield  {journal}
  {\bibinfo  {journal} {arXiv preprint arXiv:2311.13775}\ } (\bibinfo {year}
  {2023}{\natexlab{a}})}\BibitemShut {NoStop}%
\bibitem [{\citenamefont {Budinger}\ \emph {et~al.}(2024)\citenamefont
  {Budinger}, \citenamefont {Furusawa},\ and\ \citenamefont {van
  Loock}}]{budinger2024all}%
  \BibitemOpen
  \bibfield  {author} {\bibinfo {author} {\bibfnamefont {N.}~\bibnamefont
  {Budinger}}, \bibinfo {author} {\bibfnamefont {A.}~\bibnamefont {Furusawa}},\
  and\ \bibinfo {author} {\bibfnamefont {P.}~\bibnamefont {van Loock}},\
  }\bibfield  {title} {\bibinfo {title} {All-optical quantum computing using
  cubic phase gates},\ }\href@noop {} {\bibfield  {journal} {\bibinfo
  {journal} {Physical Review Research}\ }\textbf {\bibinfo {volume} {6}},\
  \bibinfo {pages} {023332} (\bibinfo {year} {2024})}\BibitemShut {NoStop}%
\bibitem [{\citenamefont {Azuma}\ \emph {et~al.}(2015)\citenamefont {Azuma},
  \citenamefont {Tamaki},\ and\ \citenamefont {Lo}}]{azuma2015all}%
  \BibitemOpen
  \bibfield  {author} {\bibinfo {author} {\bibfnamefont {K.}~\bibnamefont
  {Azuma}}, \bibinfo {author} {\bibfnamefont {K.}~\bibnamefont {Tamaki}},\ and\
  \bibinfo {author} {\bibfnamefont {H.-K.}\ \bibnamefont {Lo}},\ }\bibfield
  {title} {\bibinfo {title} {All-photonic quantum repeaters},\ }\href@noop {}
  {\bibfield  {journal} {\bibinfo  {journal} {Nature communications}\ }\textbf
  {\bibinfo {volume} {6}},\ \bibinfo {pages} {1} (\bibinfo {year}
  {2015})}\BibitemShut {NoStop}%
\bibitem [{\citenamefont {Edmonds}(1965)}]{edmonds1965paths}%
  \BibitemOpen
  \bibfield  {author} {\bibinfo {author} {\bibfnamefont {J.}~\bibnamefont
  {Edmonds}},\ }\bibfield  {title} {\bibinfo {title} {Paths, trees, and
  flowers},\ }\href@noop {} {\bibfield  {journal} {\bibinfo  {journal}
  {Canadian Journal of mathematics}\ }\textbf {\bibinfo {volume} {17}},\
  \bibinfo {pages} {449} (\bibinfo {year} {1965})}\BibitemShut {NoStop}%
\bibitem [{\citenamefont {Kolmogorov}(2009)}]{kolmogorov2009blossom}%
  \BibitemOpen
  \bibfield  {author} {\bibinfo {author} {\bibfnamefont {V.}~\bibnamefont
  {Kolmogorov}},\ }\bibfield  {title} {\bibinfo {title} {Blossom v: a new
  implementation of a minimum cost perfect matching algorithm},\ }\href@noop {}
  {\bibfield  {journal} {\bibinfo  {journal} {Mathematical Programming
  Computation}\ }\textbf {\bibinfo {volume} {1}},\ \bibinfo {pages} {43}
  (\bibinfo {year} {2009})}\BibitemShut {NoStop}%
\bibitem [{not({\natexlab{b}})}]{note4}%
  \BibitemOpen
  \href@noop {} {}\bibinfo {note} {Additionally, we simulated our scheme
  without postselection for {GKP} qubits and with postselection for
  single-photon qubits. {W}e then obtained threshold squeezing levels of 8.4~dB
  and 8.7~dB with photon loss $l_{\rm ck}=$ 0.1~\% and 1.0~\%, respectively.
  {T}his result shows the high photon loss tolerance of {GKP}
  qubits.}\BibitemShut {Stop}%
\bibitem [{\citenamefont {Fukui}(2023)}]{fukui2023high}%
  \BibitemOpen
  \bibfield  {author} {\bibinfo {author} {\bibfnamefont {K.}~\bibnamefont
  {Fukui}},\ }\bibfield  {title} {\bibinfo {title} {High-threshold
  fault-tolerant quantum computation with the gottesman-kitaev-preskill qubit
  under noise in an optical setup},\ }\href@noop {} {\bibfield  {journal}
  {\bibinfo  {journal} {Physical Review A}\ }\textbf {\bibinfo {volume}
  {107}},\ \bibinfo {pages} {052414} (\bibinfo {year} {2023})}\BibitemShut
  {NoStop}%
\bibitem [{not({\natexlab{c}})}]{note1}%
  \BibitemOpen
  \href@noop {} {}\bibinfo {note} {{T}he phenomenological error model for the
  {GKP} qubits considers the case where {GKP} qubits with infinite squeezing
  are concatenated with a higher-level quantum error-correcting code. {T}hen, a
  {G}aussian quantum channel is applied to each {GKP} qubit, and the ideal
  {GKP} qubits are replaced with those for finite squeezing. {A}s a result,
  displacement errors occurring in one {GKP} qubit do not propagate to other
  {GKP} qubits.}\BibitemShut {Stop}%
\bibitem [{not({\natexlab{d}})}]{note2}%
  \BibitemOpen
  \href@noop {} {}\bibinfo {note} {{F}or the error model, the error
  probabilities are derived from the GKP qubit itself, described in Eq.~(2),
  and an assumed depolarization error rate of $4.2\times 10^{-5}$. The reason
  that the obtained threshold is slightly higher than that of the
  phenomenological model is that we take into account the photon depolarization
  error.}\BibitemShut {Stop}%
\bibitem [{\citenamefont {Bennett}\ and\ \citenamefont
  {Brassard}(2014)}]{bennett2014quantum}%
  \BibitemOpen
  \bibfield  {author} {\bibinfo {author} {\bibfnamefont {C.~H.}\ \bibnamefont
  {Bennett}}\ and\ \bibinfo {author} {\bibfnamefont {G.}~\bibnamefont
  {Brassard}},\ }\bibfield  {title} {\bibinfo {title} {Quantum cryptography:
  Public key distribution and coin tossing},\ }\href@noop {} {\bibfield
  {journal} {\bibinfo  {journal} {Theoretical computer science}\ }\textbf
  {\bibinfo {volume} {560}},\ \bibinfo {pages} {7} (\bibinfo {year}
  {2014})}\BibitemShut {NoStop}%
\bibitem [{\citenamefont {Briegel}\ \emph {et~al.}(1998)\citenamefont
  {Briegel}, \citenamefont {D{\"u}r}, \citenamefont {Cirac},\ and\
  \citenamefont {Zoller}}]{briegel1998quantum}%
  \BibitemOpen
  \bibfield  {author} {\bibinfo {author} {\bibfnamefont {H.-J.}\ \bibnamefont
  {Briegel}}, \bibinfo {author} {\bibfnamefont {W.}~\bibnamefont {D{\"u}r}},
  \bibinfo {author} {\bibfnamefont {J.~I.}\ \bibnamefont {Cirac}},\ and\
  \bibinfo {author} {\bibfnamefont {P.}~\bibnamefont {Zoller}},\ }\bibfield
  {title} {\bibinfo {title} {Quantum repeaters: the role of imperfect local
  operations in quantum communication},\ }\href@noop {} {\bibfield  {journal}
  {\bibinfo  {journal} {Physical Review Letters}\ }\textbf {\bibinfo {volume}
  {81}},\ \bibinfo {pages} {5932} (\bibinfo {year} {1998})}\BibitemShut
  {NoStop}%
\bibitem [{\citenamefont {Kimble}(2008)}]{kimble2008quantum}%
  \BibitemOpen
  \bibfield  {author} {\bibinfo {author} {\bibfnamefont {H.~J.}\ \bibnamefont
  {Kimble}},\ }\bibfield  {title} {\bibinfo {title} {The quantum internet},\
  }\href@noop {} {\bibfield  {journal} {\bibinfo  {journal} {Nature}\ }\textbf
  {\bibinfo {volume} {453}},\ \bibinfo {pages} {1023} (\bibinfo {year}
  {2008})}\BibitemShut {NoStop}%
\bibitem [{\citenamefont {Fukui}\ \emph {et~al.}(2021)\citenamefont {Fukui},
  \citenamefont {Alexander},\ and\ \citenamefont {van Loock}}]{fukui2021all}%
  \BibitemOpen
  \bibfield  {author} {\bibinfo {author} {\bibfnamefont {K.}~\bibnamefont
  {Fukui}}, \bibinfo {author} {\bibfnamefont {R.~N.}\ \bibnamefont
  {Alexander}},\ and\ \bibinfo {author} {\bibfnamefont {P.}~\bibnamefont {van
  Loock}},\ }\bibfield  {title} {\bibinfo {title} {All-optical long-distance
  quantum communication with gottesman-kitaev-preskill qubits},\ }\href@noop {}
  {\bibfield  {journal} {\bibinfo  {journal} {Physical Review Research}\
  }\textbf {\bibinfo {volume} {3}},\ \bibinfo {pages} {033118} (\bibinfo {year}
  {2021})}\BibitemShut {NoStop}%
\bibitem [{\citenamefont {Rozp{\k{e}}dek}\ \emph {et~al.}(2021)\citenamefont
  {Rozp{\k{e}}dek}, \citenamefont {Noh}, \citenamefont {Xu}, \citenamefont
  {Guha},\ and\ \citenamefont {Jiang}}]{rozpkedek2021quantum}%
  \BibitemOpen
  \bibfield  {author} {\bibinfo {author} {\bibfnamefont {F.}~\bibnamefont
  {Rozp{\k{e}}dek}}, \bibinfo {author} {\bibfnamefont {K.}~\bibnamefont {Noh}},
  \bibinfo {author} {\bibfnamefont {Q.}~\bibnamefont {Xu}}, \bibinfo {author}
  {\bibfnamefont {S.}~\bibnamefont {Guha}},\ and\ \bibinfo {author}
  {\bibfnamefont {L.}~\bibnamefont {Jiang}},\ }\bibfield  {title} {\bibinfo
  {title} {Quantum repeaters based on concatenated bosonic and
  discrete-variable quantum codes},\ }\href@noop {} {\bibfield  {journal}
  {\bibinfo  {journal} {npj Quantum Information}\ }\textbf {\bibinfo {volume}
  {7}},\ \bibinfo {pages} {102} (\bibinfo {year} {2021})}\BibitemShut {NoStop}%
\bibitem [{\citenamefont {Rozp{\k{e}}dek}\ \emph {et~al.}(2023)\citenamefont
  {Rozp{\k{e}}dek}, \citenamefont {Seshadreesan}, \citenamefont {Polakos},
  \citenamefont {Jiang},\ and\ \citenamefont {Guha}}]{rozpkedek2023all}%
  \BibitemOpen
  \bibfield  {author} {\bibinfo {author} {\bibfnamefont {F.}~\bibnamefont
  {Rozp{\k{e}}dek}}, \bibinfo {author} {\bibfnamefont {K.~P.}\ \bibnamefont
  {Seshadreesan}}, \bibinfo {author} {\bibfnamefont {P.}~\bibnamefont
  {Polakos}}, \bibinfo {author} {\bibfnamefont {L.}~\bibnamefont {Jiang}},\
  and\ \bibinfo {author} {\bibfnamefont {S.}~\bibnamefont {Guha}},\ }\bibfield
  {title} {\bibinfo {title} {All-photonic gottesman-kitaev-preskill--qubit
  repeater using analog-information-assisted multiplexed entanglement
  ranking},\ }\href@noop {} {\bibfield  {journal} {\bibinfo  {journal}
  {Physical Review Research}\ }\textbf {\bibinfo {volume} {5}},\ \bibinfo
  {pages} {043056} (\bibinfo {year} {2023})}\BibitemShut {NoStop}%
\bibitem [{\citenamefont {Schmidt}\ \emph {et~al.}(2024)\citenamefont
  {Schmidt}, \citenamefont {Miller},\ and\ \citenamefont {van
  Loock}}]{schmidt2024error}%
  \BibitemOpen
  \bibfield  {author} {\bibinfo {author} {\bibfnamefont {F.}~\bibnamefont
  {Schmidt}}, \bibinfo {author} {\bibfnamefont {D.}~\bibnamefont {Miller}},\
  and\ \bibinfo {author} {\bibfnamefont {P.}~\bibnamefont {van Loock}},\
  }\bibfield  {title} {\bibinfo {title} {Error-corrected quantum repeaters with
  gottesman-kitaev-preskill qudits},\ }\href@noop {} {\bibfield  {journal}
  {\bibinfo  {journal} {Physical Review A}\ }\textbf {\bibinfo {volume}
  {109}},\ \bibinfo {pages} {042427} (\bibinfo {year} {2024})}\BibitemShut
  {NoStop}%
\bibitem [{\citenamefont {H{\"a}ussler}\ and\ \citenamefont {van
  Loock}(2024)}]{haussler2024quantum}%
  \BibitemOpen
  \bibfield  {author} {\bibinfo {author} {\bibfnamefont {S.}~\bibnamefont
  {H{\"a}ussler}}\ and\ \bibinfo {author} {\bibfnamefont {P.}~\bibnamefont {van
  Loock}},\ }\bibfield  {title} {\bibinfo {title} {Quantum repeaters based on
  stationary gottesman-kitaev-preskill qubits},\ }\href@noop {} {\bibfield
  {journal} {\bibinfo  {journal} {arXiv preprint arXiv:2406.07158}\ } (\bibinfo
  {year} {2024})}\BibitemShut {NoStop}%
\bibitem [{\citenamefont {Motes}\ \emph {et~al.}(2017)\citenamefont {Motes},
  \citenamefont {Baragiola}, \citenamefont {Gilchrist},\ and\ \citenamefont
  {Menicucci}}]{motes2017encoding}%
  \BibitemOpen
  \bibfield  {author} {\bibinfo {author} {\bibfnamefont {K.~R.}\ \bibnamefont
  {Motes}}, \bibinfo {author} {\bibfnamefont {B.~Q.}\ \bibnamefont
  {Baragiola}}, \bibinfo {author} {\bibfnamefont {A.}~\bibnamefont
  {Gilchrist}},\ and\ \bibinfo {author} {\bibfnamefont {N.~C.}\ \bibnamefont
  {Menicucci}},\ }\bibfield  {title} {\bibinfo {title} {Encoding qubits into
  oscillators with atomic ensembles and squeezed light},\ }\href@noop {}
  {\bibfield  {journal} {\bibinfo  {journal} {Physical Review A}\ }\textbf
  {\bibinfo {volume} {95}},\ \bibinfo {pages} {053819} (\bibinfo {year}
  {2017})}\BibitemShut {NoStop}%
\bibitem [{\citenamefont {Weigand}\ and\ \citenamefont
  {Terhal}(2018)}]{weigand2018generating}%
  \BibitemOpen
  \bibfield  {author} {\bibinfo {author} {\bibfnamefont {D.~J.}\ \bibnamefont
  {Weigand}}\ and\ \bibinfo {author} {\bibfnamefont {B.~M.}\ \bibnamefont
  {Terhal}},\ }\bibfield  {title} {\bibinfo {title} {Generating grid states
  from schr{\"o}dinger-cat states without postselection},\ }\href@noop {}
  {\bibfield  {journal} {\bibinfo  {journal} {Physical Review A}\ }\textbf
  {\bibinfo {volume} {97}},\ \bibinfo {pages} {022341} (\bibinfo {year}
  {2018})}\BibitemShut {NoStop}%
\bibitem [{\citenamefont {Arrazola}\ \emph {et~al.}(2019)\citenamefont
  {Arrazola}, \citenamefont {Bromley}, \citenamefont {Izaac}, \citenamefont
  {Myers}, \citenamefont {Br{\'a}dler},\ and\ \citenamefont
  {Killoran}}]{arrazola2019machine}%
  \BibitemOpen
  \bibfield  {author} {\bibinfo {author} {\bibfnamefont {J.~M.}\ \bibnamefont
  {Arrazola}}, \bibinfo {author} {\bibfnamefont {T.~R.}\ \bibnamefont
  {Bromley}}, \bibinfo {author} {\bibfnamefont {J.}~\bibnamefont {Izaac}},
  \bibinfo {author} {\bibfnamefont {C.~R.}\ \bibnamefont {Myers}}, \bibinfo
  {author} {\bibfnamefont {K.}~\bibnamefont {Br{\'a}dler}},\ and\ \bibinfo
  {author} {\bibfnamefont {N.}~\bibnamefont {Killoran}},\ }\bibfield  {title}
  {\bibinfo {title} {Machine learning method for state preparation and gate
  synthesis on photonic quantum computers},\ }\href@noop {} {\bibfield
  {journal} {\bibinfo  {journal} {Quantum Science and Technology}\ }\textbf
  {\bibinfo {volume} {4}},\ \bibinfo {pages} {024004} (\bibinfo {year}
  {2019})}\BibitemShut {NoStop}%
\bibitem [{\citenamefont {Eaton}\ \emph {et~al.}(2019)\citenamefont {Eaton},
  \citenamefont {Nehra},\ and\ \citenamefont {Pfister}}]{eaton2019non}%
  \BibitemOpen
  \bibfield  {author} {\bibinfo {author} {\bibfnamefont {M.}~\bibnamefont
  {Eaton}}, \bibinfo {author} {\bibfnamefont {R.}~\bibnamefont {Nehra}},\ and\
  \bibinfo {author} {\bibfnamefont {O.}~\bibnamefont {Pfister}},\ }\bibfield
  {title} {\bibinfo {title} {Non-gaussian and gottesman--kitaev--preskill state
  preparation by photon catalysis},\ }\href@noop {} {\bibfield  {journal}
  {\bibinfo  {journal} {New Journal of Physics}\ }\textbf {\bibinfo {volume}
  {21}},\ \bibinfo {pages} {113034} (\bibinfo {year} {2019})}\BibitemShut
  {NoStop}%
\bibitem [{\citenamefont {Su}\ \emph {et~al.}(2019)\citenamefont {Su},
  \citenamefont {Myers},\ and\ \citenamefont {Sabapathy}}]{su2019conversion}%
  \BibitemOpen
  \bibfield  {author} {\bibinfo {author} {\bibfnamefont {D.}~\bibnamefont
  {Su}}, \bibinfo {author} {\bibfnamefont {C.~R.}\ \bibnamefont {Myers}},\ and\
  \bibinfo {author} {\bibfnamefont {K.~K.}\ \bibnamefont {Sabapathy}},\
  }\bibfield  {title} {\bibinfo {title} {Conversion of gaussian states to
  non-gaussian states using photon-number-resolving detectors},\ }\href@noop {}
  {\bibfield  {journal} {\bibinfo  {journal} {Physical Review A}\ }\textbf
  {\bibinfo {volume} {100}},\ \bibinfo {pages} {052301} (\bibinfo {year}
  {2019})}\BibitemShut {NoStop}%
\bibitem [{\citenamefont {Fukui}\ \emph
  {et~al.}(2022{\natexlab{a}})\citenamefont {Fukui}, \citenamefont {Takeda},
  \citenamefont {Endo}, \citenamefont {Asavanant}, \citenamefont {Yoshikawa},
  \citenamefont {van Loock},\ and\ \citenamefont
  {Furusawa}}]{fukui2022efficient}%
  \BibitemOpen
  \bibfield  {author} {\bibinfo {author} {\bibfnamefont {K.}~\bibnamefont
  {Fukui}}, \bibinfo {author} {\bibfnamefont {S.}~\bibnamefont {Takeda}},
  \bibinfo {author} {\bibfnamefont {M.}~\bibnamefont {Endo}}, \bibinfo {author}
  {\bibfnamefont {W.}~\bibnamefont {Asavanant}}, \bibinfo {author}
  {\bibfnamefont {J.-i.}\ \bibnamefont {Yoshikawa}}, \bibinfo {author}
  {\bibfnamefont {P.}~\bibnamefont {van Loock}},\ and\ \bibinfo {author}
  {\bibfnamefont {A.}~\bibnamefont {Furusawa}},\ }\bibfield  {title} {\bibinfo
  {title} {Efficient backcasting search for optical quantum state synthesis},\
  }\href@noop {} {\bibfield  {journal} {\bibinfo  {journal} {Physical Review
  Letters}\ }\textbf {\bibinfo {volume} {128}},\ \bibinfo {pages} {240503}
  (\bibinfo {year} {2022}{\natexlab{a}})}\BibitemShut {NoStop}%
\bibitem [{\citenamefont {Fukui}\ \emph
  {et~al.}(2022{\natexlab{b}})\citenamefont {Fukui}, \citenamefont {Endo},
  \citenamefont {Asavanant}, \citenamefont {Sakaguchi}, \citenamefont
  {Yoshikawa},\ and\ \citenamefont {Furusawa}}]{fukui2022generating}%
  \BibitemOpen
  \bibfield  {author} {\bibinfo {author} {\bibfnamefont {K.}~\bibnamefont
  {Fukui}}, \bibinfo {author} {\bibfnamefont {M.}~\bibnamefont {Endo}},
  \bibinfo {author} {\bibfnamefont {W.}~\bibnamefont {Asavanant}}, \bibinfo
  {author} {\bibfnamefont {A.}~\bibnamefont {Sakaguchi}}, \bibinfo {author}
  {\bibfnamefont {J.-i.}\ \bibnamefont {Yoshikawa}},\ and\ \bibinfo {author}
  {\bibfnamefont {A.}~\bibnamefont {Furusawa}},\ }\bibfield  {title} {\bibinfo
  {title} {Generating the gottesman-kitaev-preskill qubit using a cross-kerr
  interaction between squeezed light and fock states in optics},\ }\href@noop
  {} {\bibfield  {journal} {\bibinfo  {journal} {Physical Review A}\ }\textbf
  {\bibinfo {volume} {105}},\ \bibinfo {pages} {022436} (\bibinfo {year}
  {2022}{\natexlab{b}})}\BibitemShut {NoStop}%
\bibitem [{\citenamefont {Hastrup}\ and\ \citenamefont
  {Andersen}(2022)}]{hastrup2022protocol}%
  \BibitemOpen
  \bibfield  {author} {\bibinfo {author} {\bibfnamefont {J.}~\bibnamefont
  {Hastrup}}\ and\ \bibinfo {author} {\bibfnamefont {U.~L.}\ \bibnamefont
  {Andersen}},\ }\bibfield  {title} {\bibinfo {title} {Protocol for generating
  optical gottesman-kitaev-preskill states with cavity qed},\ }\href@noop {}
  {\bibfield  {journal} {\bibinfo  {journal} {Physical Review Letters}\
  }\textbf {\bibinfo {volume} {128}},\ \bibinfo {pages} {170503} (\bibinfo
  {year} {2022})}\BibitemShut {NoStop}%
\bibitem [{\citenamefont {Takase}\ \emph {et~al.}(2023)\citenamefont {Takase},
  \citenamefont {Fukui}, \citenamefont {Kawasaki}, \citenamefont {Asavanant},
  \citenamefont {Endo}, \citenamefont {Yoshikawa}, \citenamefont {van Loock},\
  and\ \citenamefont {Furusawa}}]{takase2023gottesman}%
  \BibitemOpen
  \bibfield  {author} {\bibinfo {author} {\bibfnamefont {K.}~\bibnamefont
  {Takase}}, \bibinfo {author} {\bibfnamefont {K.}~\bibnamefont {Fukui}},
  \bibinfo {author} {\bibfnamefont {A.}~\bibnamefont {Kawasaki}}, \bibinfo
  {author} {\bibfnamefont {W.}~\bibnamefont {Asavanant}}, \bibinfo {author}
  {\bibfnamefont {M.}~\bibnamefont {Endo}}, \bibinfo {author} {\bibfnamefont
  {J.-i.}\ \bibnamefont {Yoshikawa}}, \bibinfo {author} {\bibfnamefont
  {P.}~\bibnamefont {van Loock}},\ and\ \bibinfo {author} {\bibfnamefont
  {A.}~\bibnamefont {Furusawa}},\ }\bibfield  {title} {\bibinfo {title}
  {Gottesman-kitaev-preskill qubit synthesizer for propagating light},\
  }\href@noop {} {\bibfield  {journal} {\bibinfo  {journal} {npj Quantum
  Information}\ }\textbf {\bibinfo {volume} {9}},\ \bibinfo {pages} {98}
  (\bibinfo {year} {2023})}\BibitemShut {NoStop}%
\bibitem [{\citenamefont {Yanagimoto}\ \emph
  {et~al.}(2023{\natexlab{b}})\citenamefont {Yanagimoto}, \citenamefont
  {Nehra}, \citenamefont {Hamerly}, \citenamefont {Ng}, \citenamefont
  {Marandi},\ and\ \citenamefont {Mabuchi}}]{yanagimoto2023quantum}%
  \BibitemOpen
  \bibfield  {author} {\bibinfo {author} {\bibfnamefont {R.}~\bibnamefont
  {Yanagimoto}}, \bibinfo {author} {\bibfnamefont {R.}~\bibnamefont {Nehra}},
  \bibinfo {author} {\bibfnamefont {R.}~\bibnamefont {Hamerly}}, \bibinfo
  {author} {\bibfnamefont {E.}~\bibnamefont {Ng}}, \bibinfo {author}
  {\bibfnamefont {A.}~\bibnamefont {Marandi}},\ and\ \bibinfo {author}
  {\bibfnamefont {H.}~\bibnamefont {Mabuchi}},\ }\bibfield  {title} {\bibinfo
  {title} {Quantum nondemolition measurements with optical parametric
  amplifiers for ultrafast universal quantum information processing},\
  }\href@noop {} {\bibfield  {journal} {\bibinfo  {journal} {PRX Quantum}\
  }\textbf {\bibinfo {volume} {4}},\ \bibinfo {pages} {010333} (\bibinfo {year}
  {2023}{\natexlab{b}})}\BibitemShut {NoStop}%
\bibitem [{\citenamefont {Pizzimenti}\ and\ \citenamefont
  {Soh}(2024)}]{pizzimenti2024optical}%
  \BibitemOpen
  \bibfield  {author} {\bibinfo {author} {\bibfnamefont {A.~J.}\ \bibnamefont
  {Pizzimenti}}\ and\ \bibinfo {author} {\bibfnamefont {D.}~\bibnamefont
  {Soh}},\ }\bibfield  {title} {\bibinfo {title} {Optical
  gottesman-kitaev-preskill qubit generation via approximate squeezed
  schr$\backslash$" odinger cat state breeding},\ }\href@noop {} {\bibfield
  {journal} {\bibinfo  {journal} {arXiv preprint arXiv:2409.06902}\ } (\bibinfo
  {year} {2024})}\BibitemShut {NoStop}%
\end{thebibliography}%

\clearpage
\onecolumngrid
\section*{Supplementary materials}
In the supplementary materials, we describe the effect of the approximated CX gate, the construction of two types of small-scale cluster states, the resource costs required for CV-FTQC, and a CV-FTQC model under the assumption of negligible photon loss.

\subsection*{A. Effect of approximated CX gate via a weak CK interaction}

\begin{figure}[!h]
\centering \includegraphics[angle=0, width=0.8\columnwidth]{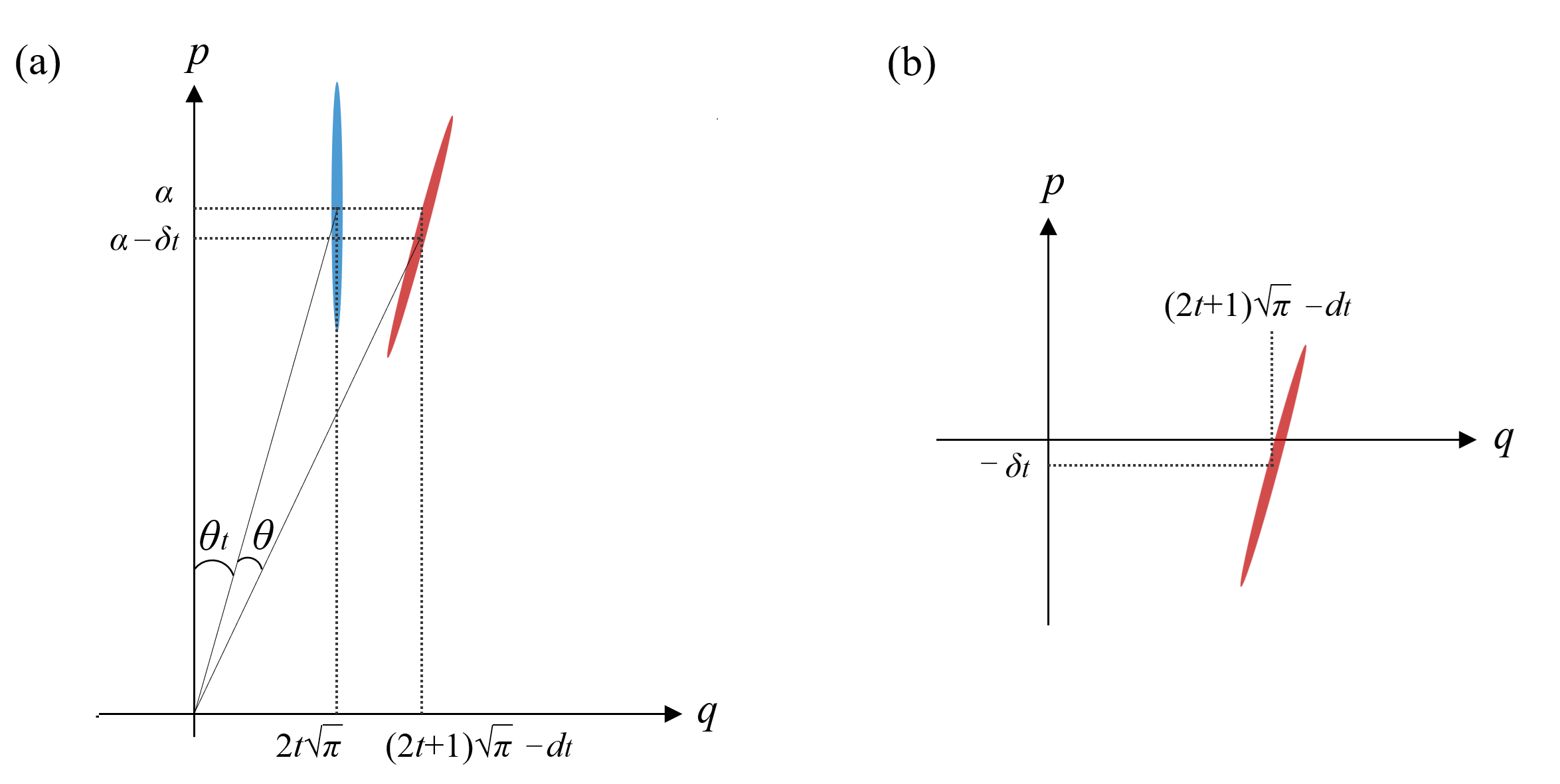} 
     \setlength\abovecaptionskip{10pt}
\caption{
The effect of the approximated CX gate via a weak CK interaction. The figure describes the approximated X gate on 
the logical 0 GKP qubit with $\delta_{t}=\alpha(1-{\rm cos}\theta)+2t\sqrt{\pi}{\rm sin}\theta$ and $d_{t}=2t\sqrt{\pi}(1-{\rm cos}\theta)$. 
(a) The state colored in red is the state after ${e}^{-\frac{i t}{\hbar}{\hat{H}_{\rm CK}}}\hat{D}(\alpha)$ on the logical 0 GKP qubit. The states colored blue and red are entangled with the single photon qubit states $\ket{0}$ and $\ket{1}$, respectively. (b) The state after $\hat{U}_{\rm CX}$. 
We note that a (Gaussian) squeezed state is considered here, rather than a (non-Gaussian) series of Gaussian peaks as for the GKP qubit,and we also have not calculated the Wigner function. Our aim here instead is to give a simple illustration and insight into the effect of the approximated CX gate.
}
\label{sfig1}
\end{figure}

We consider an imperfection in the approximated CX gate based on the CK interaction, $\hat{U}_{\rm CX}=\hat{D}(-\alpha){e}^{-\frac{i t}{\hbar}{\hat{H}_{\rm CK}}}\hat{D}(\alpha),$ where $\hat{D}(\alpha)$ is the displacement operation with the amplitude $\alpha$ in the $p$ quadrature.
Specifically, we examine the effect of the approximated CX gate on the logical 0 GKP qubit, $\ket {\widetilde{0}}$. Figure~\ref{sfig1}(a) shows the squeezed state after the operation ${e}^{-\frac{i t}{\hbar}{\hat{H}_{\rm CK}}}\hat{D}(-\alpha)$. 
After the approximated X gate, photon state $\ket{0}$ has no action on the GKP qubit, while photon state $\ket{1}$ performs the approximated displacement operation on the GKP qubit. 
Due to the approximation of the X gate, there are the imperfections in the phase rotation described by $\hat{R}(\theta)$, and displacements in both quadratures on the GKP qubit, $\hat{D}(-\delta_{t})$ and $\hat{D}(d_{t})$.
After the subsequent displacement operation $\hat{D}(-\alpha)$, $\ket {\widetilde{0}}$ entangled with $\ket{1}$ is transformed to the approximated logical 1 GKP qubit, $\ket {\widetilde{1}^{\ast}}$, as shown in Fig.~\ref{sfig1}(b).
We see that the imperfections of the approximated CX gate are negligible in our scheme, by evaluating the effects of $\hat{R}(\theta)$, $\hat{D}(-\delta_{t})$, and $\hat{D}(d_{t})$.

Here, we define the GKP qubit and the GKP qubit after the operation $\hat{U}_{\rm CX}$ as $ \ket {\widetilde{l}}$$\propto$$\sum_{t=- \infty}^{\infty} \int dx G_{t}^{l}(x) \ket{x}_q$ and $ \ket {\widetilde{l}^{\ast}}$$\propto$$\sum_{t=- \infty}^{\infty} \int dx {G^{\ast}}_{t}^{l}(x) \ket{x}_q$, where $G_{t}^{l}(q)$ and ${G^{\ast}}_{t}^{l}(q)$ represent the wavefunction of the $t$-th peak for the logical $l(=0,1)$ GKP qubit and that after the $\hat{U}_{\rm CX}$, respectively.
Then, the infidelity between $ \ket {\widetilde{l}}$ and $ \ket {\widetilde{l}^{\ast}}$ is calculated as $1-|\braket{\widetilde{l}|{\widetilde{l}^{\ast}}}|^2$.
Specifically, we focus on the infidelity between the $t$-th squeezed states corresponding to $G_{t}^{1}(q)$ and ${G^{\ast}}_{t}^{1}(q)$, which is obtained from $1-|\braket{{\rm Sq}|{\rm Sq}^{\ast}}|^2$, where $\ket{\rm Sq}$ represents a squeezed vacuum state and $\ket{{\rm Sq}^{\ast}}$ is given by $\hat{D}(d_{t})\hat{D}(-\delta_{t})\hat{R}(\theta)\ket{\rm Sq}$.
For $\delta_{t}$ and $d_{t}$, using a simple trigonometric calculation, $\delta_{t}$ and $d_{t}$ are calculated as $\delta_{t}=\alpha(1-{\rm cos}\theta)+2t\sqrt{\pi}{\rm sin}\theta$ and $d_{t}=2t\sqrt{\pi}(1-{\rm cos}\theta)$. In our scheme, we set $\alpha_0\sim 3\times 10^{5}$ and $\theta=10^{-5}$ as assumed in Refs.~\cite{nemoto2004nearly,munro2005weak}, which results in $\delta_{t}\sim10^{-6}$ and $d_{t}\sim 10^{-10}$.
Then, we find that the infidelities $1-|\braket{{\rm Sq}|{\rm Sq}^{\ast}}|^2$ with a squeezing level 10~dB for $t$=0, 5, and 10, are equal to $1.2\times10^{-9}$, $2.2\times10^{-9}$, and $4.6\times10^{-9}$, respectively.
Thus, the effect of the approximated CX gate is assumed to be negligibly small in our scheme.

\subsection*{B. Construction of the hybrid cluster states}
\begin{figure}[!h]
\centering \includegraphics[angle=0, width=1.0\columnwidth]{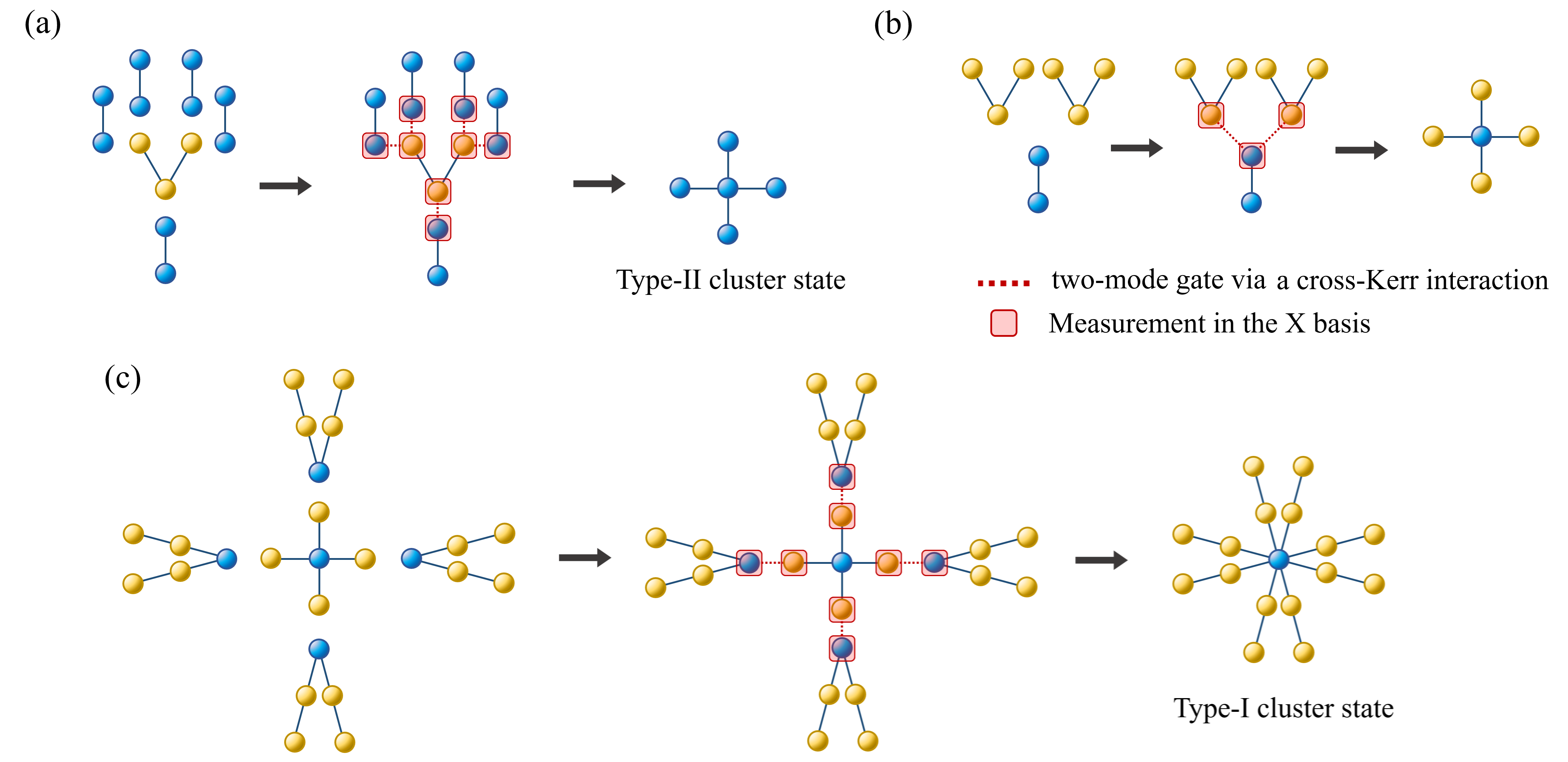} 
     \setlength\abovecaptionskip{0pt}
\caption{
Schematic view of the construction of the two types of hybrid cluster states.
(a) The preparation of the type-II cluster state.
(b) The preparation of the hybrid cluster state with four leaf photons.
(c) The preparation of the type-I cluster state.
}
\label{sfig2}
\end{figure}

We consider the construction of the two-types of small-scale cluster states described in Fig.~\ref{fig1} (c) in the main text.
We firstly prepare the elemental states, the GHZ state of photon and qunaught states.
The two qunaught states are used to prepare the Bell pair of the GKP qubits by a beam splitter coupling.
For the construction of the type-II cluster state, four Bell pairs of the GKP qubits and a single GHZ state are used to generate the type-II cluster state by using the hybrid BMs, described in Fig.~\ref{sfig2}(a). 
In the hybrid BM, we use the state for the 3D cluster state construction if all measurements for the HRM of the GKP qubit and the detection of the photon are successful.
For the construction of the type-I cluster state, we generate the hybrid cluster state with four leaf photons from the two GHZ states of photons and the GKP Bell pair by using the hybrid BMs, as described in Fig.~\ref{sfig2}(b).
Then, the type-I cluster state is constructed from the single hybrid cluster state with four leaf photons and four hybrid cluster states described in Fig.~\ref{fig1}(f) in the main text, as shown in Fig.~\ref{sfig2}(c).

\subsection*{C. Estimation of resource cost}
The resource costs for the type-I and -II cluster states, $N^{\rm I}$ and $N^{\rm II}$, are estimated from the success probabilities of the preparation of the type-I and -II cluster states, respectively.
$N^{\rm I(II)}$ is calculated by $N^{\rm I(II)}_{\rm GKP}+N^{\rm I(II)}_{\rm photon}$, where $N^{\rm I(II)}_{\rm GKP}$ and $N^{\rm I(II)}_{\rm photon}$
are the total numbers of the GKP Bell pairs and that of the GHZ state of photons for the preparation of the type-I(II) cluster state, respectively.
Then, the required number of elemental states for each node in the 3D cluster state is estimated by half the sum of the resource costs for the type-I and -II cluster states, $\frac{N^{\rm I}+N^{\rm II}}{2}$.

For the preparation of the type-I cluster, we firstly consider the required number of the elemental states for the hybrid cluster state described in Fig.~\ref{fig1}(f) in the main text. The required number of the GKP Bell pairs and that of the GHZ state of photons, $n_1$ and $n_2$, are obtained from $n_1=\left(\frac{1}{P_{\rm GKP, 2}}\right)\left(\frac{1}{P_{\rm photon, 1}}\right)^2$ and $n_2=2\left(\frac{1}{P_{\rm GKP, 2}}\right)\left(\frac{1}{P_{\rm photon, 1}}\right)^2$, respectively, where $P_{{\rm GKP},i}$ and $P_{{\rm photon},i}$ represent the success probabilities of HRM of the GKP qubit and that of a single photon detection after $i$ times approximated CX gates, respectively.
We next consider the required number of the elemental states for the hybrid cluster state described in Fig.~\ref{sfig2}(b). For this hybrid cluster state, the required number of the GKP Bell pairs and that of the GHZ states of photons are also obtained from $n_1$ and $n_2$, respectively. 
Then, the required number of the GKP Bell pairs and that of the GHZ states of photons for the type-I cluster state are obtained from $5n_1\left(\frac{1}{P_{\rm GKP, 1}P_{\rm photon, 1}}\right)^4$ and $5n_2\left(\frac{1}{P_{\rm GKP, 1}P_{\rm photon, 1}}\right)^4$, respectively.
Consequently, $N^{\rm I}$ is estimated as
\begin{equation}
N^{\rm I}=5(n_1+n_2)\left(\frac{1}{P_{\rm GKP, 1}P_{\rm photon, 1}}\right)^4=15n_1\left(\frac{1}{P_{\rm GKP, 1}P_{\rm photon, 1}}\right)^4=15\left(\frac{1}{P_{\rm GKP, 2}}\right)\left(\frac{1}{P_{\rm GKP, 1}}\right)^4\left(\frac{1}{P_{\rm photon, 1}}\right)^6.
\end{equation}
For the preparation of the type-II cluster, $N^{\rm II}_{\rm GKP}$ and $N^{\rm II}_{\rm photon}$ are calculated by $5\left(\frac{1}{P_{\rm GKP,1}}\right)^5\left(\frac{1}{P_{\rm photon, 2}}\right)^3$ and $\left(\frac{1}{P_{\rm GKP,1}}\right)^5\left(\frac{1}{P_{\rm photon, 2}}\right)^3$, respectively. 
$N^{\rm II}$ is estimated as
\begin{equation}
N^{\rm II}=6\left(\frac{1}{P_{\rm GKP,1}}\right)^5\left(\frac{1}{P_{\rm photon, 2}}\right)^3.
\end{equation}
As a result, the required number of elemental states for each node in the 3D cluster state is estimated as 
\begin{equation}
\frac{N^{\rm I}+N^{\rm II}}{2}=\frac{15}{2}\left(\frac{1}{P_{\rm GKP, 2}}\right)\left(\frac{1}{P_{\rm GKP, 1}}\right)^4\left(\frac{1}{P_{\rm photon, 1}}\right)^6+3\left(\frac{1}{P_{\rm GKP,1}}\right)^5\left(\frac{1}{P_{\rm photon, 2}}\right)^3.
\end{equation}

\subsection*{D. CV-FTQC model under the assumption of negligible photon loss}

\begin{figure}[!h]
\centering \includegraphics[angle=0, width=0.5\columnwidth]{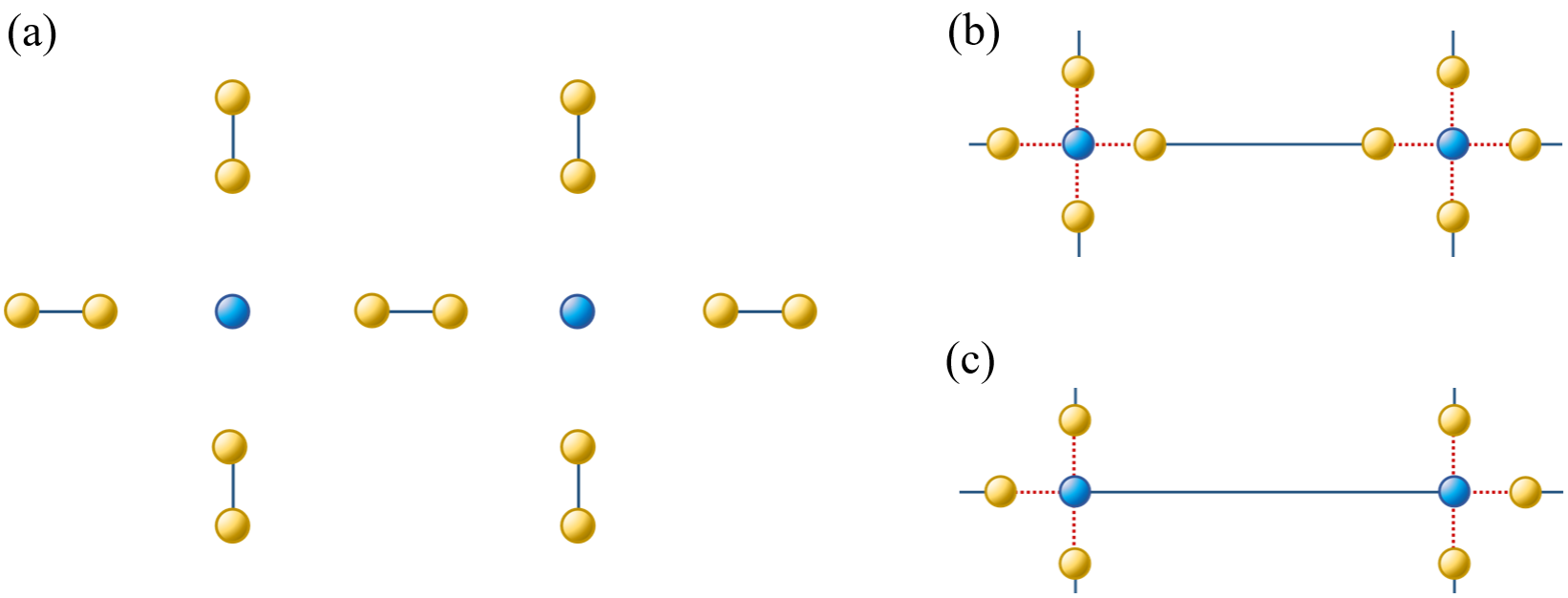} 
     \setlength\abovecaptionskip{10pt}
\caption{Schematic for the construction of a large-scale cluster state under very low photon loss. (a) The GKP qubits are used for a node qubit of the 3D cluster state described in Fig.1 (e) in the main text. The entangled photons are used for the entanglement generation between node qubits. (b) The entanglement generation between a photon and the GKP qubit via the CK interaction. (c) The entanglement generation between neighboring GKP qubits by the measurements of photons in the $X$ basis.
We note that the entanglement generation is implemented deterministically due to very low photon loss. 
}
\label{sfig3}
\end{figure}

We consider a CV-FTQC model as shown in Fig.~\ref{sfig3}: (1) Deterministic preparation of elementary quantum states, e.g., the GKP qubit and the Bell states of photons. (2) Deterministic hybrid BM, where the hybrid BM can be implemented deterministically because the failure of single-photon detection can be neglected due to very low photon loss. (3) Deterministic construction of a large-scale cluster state, also thanks to very low photon loss.
After preparing the elementary quantum states, one photon from the Bell pair is sent to the neighboring node.
Then, we implement the hybrid BM via the weak CK scheme to entangle the neighboring GKP qubits.
Due to the deterministic hybrid BM, a large-scale cluster state can be obtained deterministically.

\end{document}